\documentclass[a4,20pt]{article}

\usepackage[dvips]{graphicx}
\usepackage{amsmath,amsthm,amssymb}

\newcommand{\bP}{\mathbb{P}}

\newcommand{\bC}{\mathbb{C}}
\newcommand{\bZ}{\mathbb{Z}}
\newcommand{\bF}{\mathbb{F}}

\begin{document}

\large

\begin{titlepage}
\vspace*{5mm}
\hfill
\vbox{
    \halign{#\hfil         \cr
            YITP-05-47 \cr
           } 
      }  
\vspace*{10mm}

\Large

\begin{center}
{\LARGE {\bf \LARGE String Theories on Flat Supermanifolds \\
}}
\vspace*{15mm}
{{\sc Tatsuya Tokunaga}
\footnote{ \large e-mail: {\tt tokunaga@yukawa.kyoto-u.ac.jp}}}

\vspace*{7mm}
{\it 
Yukawa Institute for Theoretical Physics, \\ 
Kyoto University, Kyoto 606-8502, Japan 
} \\

\end{center}

\large 

\vspace*{1.5cm}

\begin{abstract}
\large 
~ \\
We construct bosonic string theories, RNS string theories and heterotic string theories on flat supermanifolds.  
For these string theories, we show cancellations of the central charges and modular invariance.  
Bosonic string theories on supermanifolds have dimensions $(D_B,D_F)=(26,0),(28,2),(30,4),\cdots$, 
where $D_B$ and $D_F$ are the numbers of bosonic coordinates and fermionic coordinates, respectively.  
We show that in type II string theories the one loop vacuum amplitudes vanish.  From this result, we can suggest the existence of supersymmetry on supermanifolds.  
As examples of the heterotic string theories, we construct those whose massless spectra are related to $N=1$ supergravity theories 
and $N=1$ super Yang-Mills theories with orthosymplectic supergroups on the bosonic flat 10 dimensional Minkowski space.  
Also, we construct D-branes on supermanifolds and compute tensions of the D-branes.  
We show that the number of fermionic coordinates contributes to the tensions of the D-branes as an inverse power of the contribution of bosonic coordinates.  
Moreover, we find some configurations of two D-branes which satisfy the BPS-like no-force conditions if $\nu_B - \nu_F = 0,4$ and $8$, 
where $\nu_B$ and $\nu_F$ are the numbers of Dirichlet-Neumann directions in the bosonic coordinates and in the fermionic coordinates, respectively.
\end{abstract}

\vskip 2cm


\end{titlepage}
\newpage

\large

\section{\Large Introduction}

Recently Witten introduced twistor string theory as a B-twisted topological string theory on a supermanifold, supertwistor space ${\bC \bP}^{3|4}$\cite{1}.  
Topological string theories on supermanifolds are generally defined by twisting the non-linear sigma models on K\"{a}hler supermanifolds, 
though the K\"{a}hler supermanifolds should be Calabi-Yau supermanifolds so that the topological B models are well-defined.  
Twistor string theory is related to the conformal supergravity and super Yang-Mills theory in four dimensional space by use of the Penrose transformation \cite{1} \cite{5}.  
Also, twistor string theory suggests a powerful prescription for computing amplitudes of Yang-Mills theory concretely.  
Moreover, topological string theories on other supermanifolds such as toric supermanifolds were introduced 
and the mirror symmetry between supermanifolds were studied \cite{1.5} \cite{2} \cite{a2}.  
Thus, it is interesting to study string theories on supermanifolds and to find a beautiful structure beyond the case of bosonic manifolds.

Also, it is well-known that the partition functions of the topological string theories on bosonic manifolds are equivalent 
to the topological amplitudes of the type II superstring theory \cite{3.5} \cite{3}.  
Topological amplitudes are the specific amplitudes of type II superstring theory with anti-selfdual gravitons and anti-selfdual graviphotons.  
Therefore, it would be natural to ask whether we can construct string theories on supermanifolds which are related to topological string theories on supermanifolds.  
Here, as a first step we construct string theories on flat supermanifolds by following the techniques of usual string theories on a flat Minkowski bosonic manifold \cite{11} \cite{12}.  
Though it would also be very interesting to check whether the topological string theories on supermanifolds are related to the type II string theories on supermanifolds, which will be constructed later, 
we do not discuss it in this paper.

\begin{figure}[htbp]
  \begin{center}
    \includegraphics[width=0.55\textwidth]{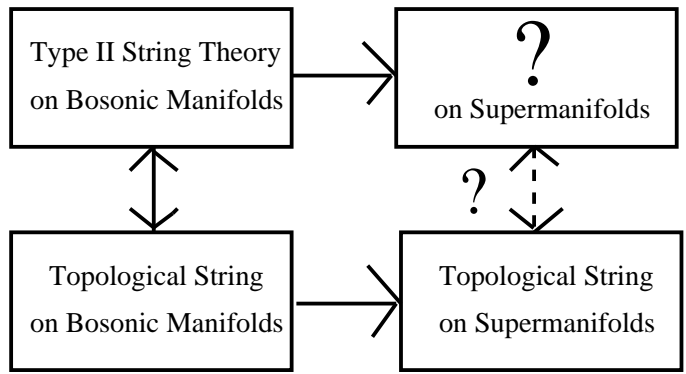}
  \end{center}
\caption{The relation between type II string theory and topological string theory on bosonic manifolds and supermanifolds.}
\label{fig:top-II.eps}
 \end{figure}

In this paper, we study string theories on flat supermanifolds.  
We construct bosonic string theories, type II, type 0, type I string theories and heterotic string theories on flat supermanifolds.  
For these string theories, we show cancellations of the central charges and modular invariance.  
We find that the bosonic string theories on supermanifolds have the dimensions $(D_B,D_F)=$$(26,0)$,$(28,2)$,$(30,4),\cdots$, 
where $D_B$ and $D_F$ are the numbers of bosonic coordinates and fermionic coordinates, respectively.  
For type II, type I and heterotic string theories on supermanifolds, we show that the one loop vacuum amplitudes vanish.  
This may suggest supersymmetry on supermanifolds.  
Moreover, we can construct many kinds of heterotic string theories on supermanifolds, which are modular invariant and have vanishing one loop vacuum amplitudes.  
As examples, we construct the heterotic string theories whose massless spectra are related to $N=1$ supergravity theory 
and $N=1$ super Yang-Mills theory with supergroups $Osp(2 D^L_B-20| 2 D^L_F)$ on the bosonic flat 10 dimensional Minkowski space.  

\begin{figure}[htbp]
  \begin{center}
    \includegraphics[width=0.40\textwidth]{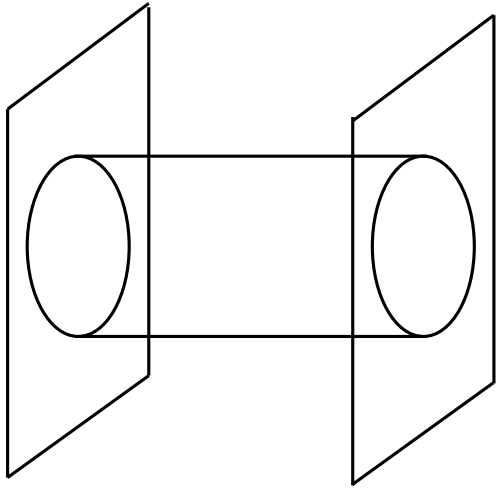}
  \end{center}
\caption{A one loop vacuum amplitude of an open string with one end on each D-brane in the open string channel. 
Also this is an exchange of a closed string between two D-branes in the closed string channel.}
\label{fig:openclosed.eps}
 \end{figure}

In \cite{01}, Polchinski showed that D-branes with RR charges have to exist as non-perturbative objects in string theory.  
The D-branes are constructed by imposing the Dirichlet boundary conditions and the endpoints of open strings can be attached to the D-branes.  
See \cite{02} for reviews.  
The force between two branes was computed both in the open string channel and in the closed string channel, 
and the BPS no-force configurations of two D-branes were found \cite{300} \cite{01} \cite{9}.  
%
Also, recently Witten introduced D1-branes in twistor string theory \cite{1}, 
and also D-branes in topological string theory on other supermanifolds were studied in \cite{2}.  
In this paper, we construct D-branes in the bosonic string theories and in the type II string theories on supermanifolds.  
The forces between two D-branes are computed both in the open string channel and in the closed string channel (Fig.\ref{fig:openclosed.eps}).  
From these results, we can obtain tensions of the D-branes and we show that the number of fermionic coordinates contributes to the tensions of the D-branes as an inverse power of the contribution of bosonic coordinates to the tensions.  
In the type II string theories, we find some configurations of two D-branes which satisfy the BPS-like no-force conditions if $\nu_B - \nu_F =0,4$ and $8$.  
Here $\nu_B$ and $\nu_F$ are the numbers of Dirichlet-Neumann directions in the bosonic coordinates and in the fermionic coordinates, respectively.  


The organization of this paper is as follows.  In section 2, bosonic string theories on flat supermanifolds are introduced.  
We show cancellations of the central charges and modular invariance.
In section 3, we consider RNS string theories and heterotic string theories on supermanifolds.  
We study the GSO projections in order to construct type II theories and type 0 theories on supermanifolds.  
Moreover, we study type I theories and heterotic string theories.  
In section 4, we construct the D-branes in the bosonic string theories and in the type II string theories on supermanifolds.  
Boundary states are constructed, and the annulus amplitudes are computed both in the open string channel and in the closed string channel to determine tensions of D-branes.  
Section 5 includes conclusion and discussion.  
We mainly obey the notations of the Polchinski's books \cite{12}.  
~ \\
~ \\
%

\section{\Large Bosonic String Theories on Flat Supermanifolds}

We construct bosonic string theories on flat supermanifolds in the same way as bosonic string theory on the flat bosonic manifold \cite{11} \cite{12}.  
For the definition of supermanifolds, we refer to \cite{4}, for example.  
As in the papers \cite{1} \cite{1.5} \cite{2}, the actions of the string theories on supermanifolds include 
the bosonic coordinates $X^{\mu}$ and the fermionic coordinates $\Theta^A$.  
Here the labels $\mu$ and $A$ run from 0 to $D_B-1$ and from 1 to $D_F$, respectively, where $D_F$ is an even number.  
We denote the dimension of the bosonic directions by $D_B$ and the dimension of the fermionic directions by $D_F$.  
Later, these dimensions will be determined by the cancellations of the central charges.
The Nambu-Goto action on supermanifolds is 
\begin{equation}
S = -  \frac{1}{2 \pi {\alpha}'} \int_M d \tau d \sigma \sqrt{ - {\rm det} \left( \partial_a {\Phi}^{P} \partial_b {\Phi}^{Q} G_{Q P} \right)} ,
\end{equation}
where $M$ is the worldsheet, $\tau$ and $\sigma$ are the worldsheet coordinates, and the indices $a,b = \tau, \sigma$.  
The coordinates $\Phi^P$ are $(X^{\mu}, \Theta^A)$, and the labels $P$ and $Q$ run over $(\mu,A)$.  
The metric on supermanifolds is denoted by $G_{PQ}$.  
The indices of the fields $\Phi^P$ are raised and lowered with the metric $G_{PQ}$ as follows;
\begin{eqnarray}
& & \Phi_P = \Phi^Q ~ G_{Q P} =  (-1)^{|P||Q|} ~ G_{P Q}  ~ \Phi^Q \nonumber, \\
& & \Phi^P = G^{P Q} ~ \Phi_Q = (-1)^{|P||Q|} ~ \Phi_Q ~ G^{ Q P } ,
\end{eqnarray}
where $|P|=0$ for the bosonic coordinates and $|P|=1$ for the fermionic coordinates.

The Polyakov action on flat supermanifolds is 
\begin{equation}
S = \frac{1}{2 \pi{{\alpha}'} } \int d^2 z  \left(   \partial X^{\mu} \bar{\partial} X_{\mu} 
- i   \partial \Theta^A \bar{\partial} \Theta^B G_{BA} \right)  \label{1}.
\end{equation}
Here we denote the Minkowski metric of the bosonic directions by $\eta^{\mu \nu}$ and the metric of the fermionic directions by $G_{AB}$, 
which is defined by $G_{A A-1} = - G_{A-1 A} =1 $ for all even numbers $A$ and the others are zero.  
For example, $G_{21}= - G_{12}=1$ and $G^{21}= -  G^{12} =1$.
From the equations of motion, we can write the mode expansions for open strings.  
\begin{eqnarray}
& & X^{\mu} = x^{\mu} -i {\alpha}' p^{\mu} \log |z|^2 + i \left( \frac{{\alpha}'}{2} \right)^{\frac{1}{2} } \sum_{m \not= 0 } \frac{{\alpha^\mu_m}}{m} 
\left( z^{-m} +  {\bar{z}}^{-m} \right)  ,\\
& & \Theta^A = \theta^A - i {\alpha}' w^A \log |z|^2 + i \left( \frac{{\alpha}'}{2} \right)^{\frac{1}{2} } \sum_{m \not= 0 } \frac{{\beta^A_m}}{m} 
\left( z^{-m} +  {\bar{z}}^{-m} \right)  .
\end{eqnarray}
%
%
The zero modes are written as $\alpha_0^{\mu} = \sqrt{2 {\alpha}'} p^{\mu}$ and $\beta_0^{A} = \sqrt{2 {\alpha}'} w^A $.  
Also, the operator product expansions for the coordinates $X^{\mu}$ and $\Theta^A$ are determined as 
\begin{eqnarray}
X^{\mu} (z_1) X^{\nu} (z_2) &\sim& -\frac{{\alpha}'}{2} \eta^{\mu \nu} \log | z_{12}|^2 ,\label{2.1}\\
\Theta^A (z_1) \Theta^B (z_2) &\sim& - ~ i \frac{{\alpha}'}{2} G^{AB} \log | z_{12}|^2 \label{2}.
\end{eqnarray}
From these operator product expansions, we find the commutators
\begin{eqnarray}
\left[ x^{\mu} , p^{\nu} \right] &=& i ~ \eta^{\mu \nu} ,\\
\{ \theta^A , w^B \} &=&  - ~ G^{AB} ,\\
\left[ \alpha^{\mu}_m, \alpha^{\nu}_n \right] &=& m ~  \eta^{\mu \nu} \delta_{m,-n} ,\\
\{ \beta^A_m , \beta^B_n \} &=& i ~ m ~ G^{AB} \delta_{m,-n} \label{4}.
\end{eqnarray}
The energy-momentum tensor is defined as the variation of the action (\ref{1}) with respect to the worldsheet metric.  The holomorphic part $T(z)= T_{z z}(z)$ is
\begin{equation} 
T(z) = - \frac{1}{{\alpha}'}  \partial X^{\mu} \partial  X_{\mu}  
     +   \frac{i}{{\alpha}'} ~  \partial \Theta^{A} \partial  {\Theta}^{B} G_{BA} \label{3}.
\end{equation}
Here the central charges for these conformal field theories can be computed by use of the operator product expansions (\ref{2.1}) (\ref{2}) and the energy-momentum tensor (\ref{3}).  
We compute only the fermionic parts since the computation for the bosonic parts is well-known \cite{12}.  
\begin{eqnarray}
T(z) T(z') &=&    \frac{i}{{\alpha}'} ~ : \partial \Theta^{A} \partial  {\Theta}^{B} G_{BA}: 
               ~~   \frac{i}{{\alpha}'} ~ : \partial \Theta^{C} \partial  {\Theta}^{D} G_{DC}:  \nonumber  \\
   &=& 
\exp \left(- i \frac{{\alpha}'}{2} G^{EF} \int d^2 z_1 d^2 z_2 \log | z_{12}|^2 \frac{\delta}{\delta \Theta^F} \frac{\delta}{\delta \Theta^E} \right) \nonumber \\
& & \times \exp \left(- i \frac{{\alpha}'}{2} G^{IJ} \int d^2 z_3 d^2 z_4 \log | z_{34}|^2 \frac{\delta}{\delta \Theta^J} \frac{\delta}{\delta \Theta^I} \right) \nonumber \\
& & \times \left(   \frac{i}{{\alpha}'} \right)^2 : \partial \Theta^{A} \partial  {\Theta}^{B} G_{BA}: ~ : \partial \Theta^{C} \partial  {\Theta}^{D} G_{DC}: \nonumber 
\end{eqnarray}
~~~
\begin{eqnarray}
&\sim& \frac{1}{4 (z-z')^4} G^{EF} G^{IJ} G_{BA} G_{DC} \nonumber \\ 
& & \times \left[(-{G_I}^D) (- {G_J}^B) (- {G_E}^C) {G_F}^A + {G_I}^C (- {G_J}^B) (- {G_E}^D) {G_F}^A \right] \nonumber \\
& & + \frac{4}{2(z-z')^2} \left( \frac{i}{{\alpha}'} \partial \Theta^B \partial \Theta^D \left[ G^{EF} G_{BA} G_{DC} {G_E}^C {G_F}^A \right] \right)\nonumber  \\
& & + \frac{2}{2 (z-z')} \partial \left( \frac{i}{{\alpha}'} \partial \Theta^B \partial \Theta^D \left[ G^{EF} G_{BA} G_{DC} {G_E}^C {G_F}^A \right] \right) \nonumber \\ 
&\sim& \frac{{G^A}_A}{2 (z-z')^4}  + \frac{2}{(z-z')^2} T(z') + \frac{1}{ (z-z')} \partial T(z') .
\end{eqnarray}
Therefore we can see that the central charges of the fermionic parts are ${G^A}_A = - D_F$.  
Also the ghost parts are the same as in the case of usual string theories on the flat bosonic manifold.  
After summing up the bosonic parts, the fermionic parts and the ghost parts, the central charges of these conformal field theories are $D_B - D_F- 26$.  
The conditions that the central charges of these theories vanish are 
\begin{equation}
D_B - D_F =  26. \label{a1}
\end{equation}
This equation determines the critical dimensions $(D_B, D_F)$. For example, $(D_B, D_F)=(26,0),(28,2),(30,4),\cdots$.
The Virasoro generators $L_n$ for nonzero integers $n$ are defined by 
\begin{eqnarray}
L_n &=& \int \frac{d z}{2 \pi i } z^{n+1} T(z) \nonumber \\ 
    &=& \frac{1}{2} \sum_{m} \alpha^{\mu}_{n-m} \alpha_{\mu m}
      -  \frac{i}{2} \sum_{m} \beta^{A}_{n-m} \beta_{A m}  .
\end{eqnarray}
The Virasoro generator $L_0$ is 
\begin{equation} 
L_0 = {\alpha}' p^2 + \sum_{m=1}^{\infty} \alpha^{\mu}_{-m} \alpha_{\mu m}  
   -i {\alpha}'   w^A w_A -i ~ \sum_{m=1}^{\infty} \beta^{A}_{-m} \beta_{A m}  -1 \label{5}.
\end{equation}
%
The zero point energy $-1$ is the same as the bosonic string on the flat bosonic manifold because the contributions from extra bosonic coordinates and those from fermionic coordinates are canceled.  
We can see that these Virasoro generators satisfy the Virasoro algebra.  
Also, it is easy to study the closed strings of the bosonic string theories on supermanifolds similarly.

Next we study the physical states by using the method of the old covariant quantization.  
Firstly the spectra of open strings, especially the ground states and the massless states, are discussed.  
In this paper, we call $m$ as 'mass', where $m$ is defined by the on-shell condition;
\begin{equation}
p^2 - i w^A w_A + m^2=0 .
\end{equation}
The ground state $| 0;p,w \rangle$ depends on the momenta $p^{\mu}$ and $w^A$ on supermanifolds.  
The ground state obeys the Virasoro condition: 
\begin{equation}
L_0 | 0;p,w \rangle = \left( {\alpha}' p^2 -i {\alpha}'   w^A w_A ~ -1 \right)| 0;p,w \rangle =0 \label{6}.
\end{equation}
This equation means that the ground state is a tachyonic state.  
Also we can construct the vertex operators $V_{\rm tachyon}(p,w)$ for the tachyonic ground state.
\begin{equation}
V_{\rm tachyon} = e^{i p^{\mu} X_{\mu}} e^{w^A \Theta_A} .
\end{equation}
Next the massless states $|e,f;p,w \rangle$ are given by
\begin{equation}
|e,f;p,w \rangle = \left( e^{\mu} \alpha_{-1 \mu} + f^A \beta_{-1 A} \right) |0;p,w \rangle .
\end{equation}
Here we use the Grassmann even coefficients $e^{\mu}$ and the Grassmann odd coefficients $f^A$.  
The Virasoro conditions for the massless states are 
\begin{eqnarray}
L_0 |e,f;p,w \rangle &=& \left( {\alpha}' p^2 -i {\alpha}' w^A w_A  \right) |e,f;p,w \rangle = 0  ,\\
L_1 |e,f;p,w \rangle &=& \left( 2 {\alpha}' \right)^2  \left( e^{\mu} p_{\mu} + f^A w_{A} \right) |0;p,w \rangle =0  ,\\
|e,f;p,w \rangle &\sim& |e,f;p,w \rangle + L_{-1} |0;p,w \rangle \nonumber \\
  &=&  |e,f;p,w \rangle  + \left( 2 {\alpha}' \right)^2 \left( p^{\mu} \alpha_{\mu -1}
      -i  w^{A} \beta_{A -1} \right) |0;p,w \rangle                              .
\end{eqnarray}
These equations mean that massless states are gauge fields on supermanifolds.  
Also we can construct the vertex operator $V_{\rm gauge}$ corresponding to the massless states.
\begin{equation}
V_{\rm gauge} = \left( e^{\mu} \partial X_{\mu} + f^{A} \partial \Theta_A \right) e^{i p^{\mu} X_{\mu}} e^{w^A \Theta_A} .
\end{equation}
Similarly we can study the spectra of closed strings, which include tachyonic ground states and massless states.  
The massless states are a graviton, an antisymmetric two-form field and a dilaton on supermanifolds.  
One can see that the effective actions for the bosonic string theories on supermanifolds include Yang-Mills theory and gravity theory on supermanifolds 
by use of the cubic string field theory and the $\beta$-function in a curved background.  
The actions can be written simply by adding fermionic coordinates to the Yang-Mills theory and gravity theory on bosonic manifolds.  
For example, the action of the gravity theory on supermanifolds is 
\begin{equation}
S = \int d^{D_B} x ~ d^{D_F} \theta \sqrt{- G} ~ R,
\end{equation}
where $G$ is the superdeterminant of the metric $G_{PQ}$ on supermanifolds and $R$ is the Ricci scalar for the metric $G_{PQ}$ \cite{41}.  

We discuss the norms of the physical states.  
Here we simply denote the tachyonic ground states by $|0 \rangle$ whose momenta are omitted.  
For example, we consider the excitations of $\beta^A_{-n}$ for all non-negative integers $n$ and $A=1,2$.  
From (\ref{4}), the operators $\beta^A_{-n}$ anticommutate with each other.  So, four states for a non-negative integer $n$ are 
\begin{equation}
| 0 \rangle , ~~    \frac{1}{\sqrt{2}} \left( i \beta^1_{-n}  + \beta^2_{-n} \right) |0 \rangle ,~~  
\frac{1}{\sqrt{2}} \left( i \beta^1_{-n}  - \beta^2_{-n} \right)| 0 \rangle, ~~  i \beta^1_{-n} \beta^2_{-n} | 0 \rangle.
\end{equation}
The norms for these states are 
\begin{eqnarray} 
\frac{1}{2} \langle 0 | \left( -i \beta^1_{n}  + \beta^2_{n} \right)    \left( i \beta^1_{-n}  + \beta^2_{-n} \right) |0 \rangle 
&=&   - n  \langle 0 |0 \rangle ,\\
\frac{1}{2} \langle 0 | \left( -i \beta^1_{n}  - \beta^2_{n} \right)    \left( i \beta^1_{-n}  - \beta^2_{-n} \right) |0 \rangle 
&=&   n  \langle 0 |0 \rangle ,\\
\langle 0 | \left( \beta^2_{n} \beta^1_{n} \beta^1_{-n} \beta^2_{-n} \right) | 0 \rangle 
&=&  - n^2 \langle 0 |0 \rangle .
\end{eqnarray}
Therefore, just half of the physical states have negative norms and the other physical states have positive norms.  
From this result, we can see that the string theories on supermanifolds are non-unitary theories.  

Finally we study the one loop vacuum amplitudes of the oriented closed strings for these string theories.  
The moduli parameters on the torus are denoted by $\tau = \tau_1 + i \tau_2 $ and $q$ is defined as $\exp( 2 \pi i  \tau)$.  
The torus amplitudes $Z_{T_2}$ are given by the Coleman-Weinberg formula;
\begin{equation}
Z_{T^2} =  \int_{F_0} \frac{d^2 \tau}{4 \tau_2} \int \frac{d^{D_B} p}{ (2 \pi)^{D_B} } (-  i 2 \pi)^{\frac{D_F}{2}}  \int d^{D_F} w \sum_{i \in H^{\perp}} (-1)^{{\bF}_i} 
q^{\frac{{\alpha}'}{4} (p^2 -i w^2 + m_i^2) } {\bar{q}}^{\frac{{\alpha}'}{4} (p^2 -i w^2 +{{m}}_i^2) } \label{1234}.
\end{equation}
Here ${F_0}$ is the fundamental region and ${\bF}_i$ is the number of the space-time Grassmann odd fields.  
$H^{\perp}$ is the closed string Hilbert space excluding the ghosts, the $\mu=0,1$ oscillators, and the noncompact momenta.  
The masses $m_i$ are the eigenvalues of the Hamiltonian $H$ for the Hilbert space $H^{\perp}$:
\begin{equation}
H = \sum_{m=1}^{\infty} \alpha^{j}_{-m} \alpha_{j m} - i  \sum_{m=1}^{\infty} \beta^{A}_{-m} \beta_{A m} -1 .
\end{equation}
Here $j = 2,3,\cdots,D_B-1$.  The bosonic coordinate part $Z_{X} (\tau)$ of the partition functions is 
\begin{eqnarray}
Z_{X} (\tau) 
 &=& \left( q \bar{q} \right)^{- \frac{1}{24}} \prod_{\mu, n} \sum^{\infty}_{N_{\mu n}, {\tilde{N}}_{\mu n} =0 } q^{n N_{\mu n}} {\bar{q}}^{n {\tilde{N}}_{\mu n}} \nonumber \\
  &=&  \left( q \bar{q} \right)^{- \frac{1}{24}} \left( \prod^{\infty}_{n=1} \left( 1- q^n \right)^{-1} \left( 1- {\bar{q}}^n \right)^{-1} \right)  \nonumber \\
&=& |\eta(\tau)|^{- 2 }.
\end{eqnarray}
The fermionic coordinate part $Z_{\Theta} (\tau)$ is 
\begin{eqnarray}
Z_{\Theta} (\tau) &=& \left( q \bar{q} \right)^{+ \frac{1}{24}} \prod_{A, n} 
\sum^{1}_{N_{A n}, {\tilde{N}}_{A n} =0} (-q)^{n N_{A n}} (-{\bar{q}})^{n {\tilde{N}}_{A n}} \nonumber \\
  &=& \left( q \bar{q} \right)^{ \frac{1}{24}} 
\left( \prod^{\infty}_{n=1} \left( 1- q^n \right) \left( 1- {\bar{q}}^n \right) \right) \nonumber \\
&=& |\eta(\tau)|^{2}.
\end{eqnarray}
After integrating over the momenta, the one loop vacuum amplitudes turns out to be
\begin{eqnarray}
Z_{T^2} &=& \int_{F_0} \frac{d \tau d {\bar{\tau}}}{4 \tau_2} 
i \left( 4 {\pi}^2 {\alpha}' \tau_2 \right)^{- \frac{D_B - D_F}{2} } 
Z_{X}^{D_B-2}(\tau) Z_{\Theta}^{D_F} (\tau) \nonumber \\
   &=& i \left( 4 {\pi}^2 {\alpha}'  \right)^{- \frac{D_B - D_F}{2} } \int_{F_0}
\frac{ d \tau d {\bar{\tau}}}{4 \tau_2^2} \left( \tau_2 | \eta(\tau) |^4 \right)^{- \frac{D_B-D_F-2}{2}} .
\end{eqnarray}
From these results, we can see that the modular parameter dependent parts of the one loop vacuum amplitudes are completely determined by $D_B-D_F$, 
which has to be 26 from (\ref{a1}).  
In conclusion, the bosonic string theories on supermanifolds with vanishing central charges are modular invariant.

Anywhere in this paper, we compute amplitudes per unit volume.  
All string amplitudes are proportional to $V_B V_F $, 
where $V_B$ and $V_F$ are the volume of the bosonic directions and the fermionic directions, respectively.  
Usually, $V_B$ is regularized if we consider a compact space instead of the non-compact space.  
We expect that $V_F$ can be regularized to a non-zero finite volume, if we use the noncommutativity of the fermionic coordinates 
and the heat kernel method, for example.  
%
%
%

Also, we can study open string similarly.  
For open string, we introduce the Chan-Paton factor, which induces non Abelian gauge groups.  
When the open string and the unoriented closed string are considered, 
the non-zero one loop vacuum amplitude shows that a tadpole exists.  
In the case of the open string theory on supermanifolds, one can compute the one loop vacuum amplitudes similarly, 
and these amplitudes are mostly equal to the one loop vacuum amplitudes in the case of the bosonic manifold.  
Therefore, we can see that the tadpole is canceled as usual if the gauge group is $SO(2^{13})$.  
~ \\
~ \\
~ \\

\section{\Large RNS String Theories and Heterotic String Theories on Flat Supermanifolds}

It is well-known that there are five superstring theories on 10 dimensional bosonic manifolds; type II AB, type I and heterotic $SO(32)$ and $E_8 \times E_8$ superstring theories.  
Here we construct the type II, type I and heterotic string theories on supermanifolds.  
We show the physical spectra, the cancellations of central charges and modular invariance.

\subsection{\Large Type II String Theories on Flat Supermanifolds}

In order to construct RNS string theories on supermanifolds, superpartners on the worldsheet are introduced.  
The superpartners of the fields $X^{\mu}$ are denoted by $\psi^{\mu}$ and ${\tilde{\psi}}^\mu$.  
The superpartners of the fields $\Theta^A$ are denoted by $\rho^A $ and ${\tilde{\rho}}^A $.  
The actions for the RNS string theories are written as
\begin{eqnarray}
S &=& \frac{1}{4 \pi} \int d^2 z  \left( \frac{2}{{\alpha}'}  \partial X^{\mu} \bar{\partial} X_{\mu} 
- i \frac{2}{{\alpha}'}  \partial \Theta^A \bar{\partial} \Theta_A \right.\nonumber \\
    & & \left. ~~~~ +  \psi^{\mu} \bar{\partial} \psi_{\mu} +  {\tilde{\psi}}^{\mu} \partial {\tilde{\psi}}_\mu 
             - i  \rho^A \bar{\partial} \rho_A  - i  {\tilde{\rho}}^A \partial {\tilde{\rho}}_A  \right).
\end{eqnarray}
From this action, the operator product expansions are 
\begin{eqnarray}
\psi^{\mu} ( z) \psi^{\nu} (0) &\sim& \frac{\eta^{\mu \nu}}{z} , \\
{\tilde{\psi}}^{\mu} ( \bar{z}) {\tilde{\psi}}^{\mu} (0) &\sim& \frac{\eta^{\mu \nu}}{\bar{z}} ,\\
\rho^A (z) \rho^B (0) &\sim& \frac{i ~ G^{AB}}{z} ,\\
{\tilde{\rho}}^A (\bar{z}) {\tilde{\rho}}^B (0)  &\sim& \frac{i ~ G^{AB}}{\bar{z}} .
\end{eqnarray}
The mode expansions of these fields are 
\begin{eqnarray}
\psi^{\mu} ( z) = \sum_{r \in \bZ + \nu } \frac{ \psi^{\mu}_{r} }{z^{r + \frac{1}{2}}} , ~~ & & ~~
{\tilde{\psi}}^{\mu} ( \bar{z})  = \sum_{r \in \bZ + \tilde{\nu} } \frac{ {\tilde{\psi}}^{\mu}_{r} }{{\bar{z}}^{r + \frac{1}{2}}} , \\
\rho^A (z) = \sum_{r \in \bZ + \nu } \frac{ \rho^{A}_{r} }{z^{r + \frac{1}{2}}} , ~~ & & ~~ 
{\tilde{\rho}}^A (\bar{z}) = \sum_{r \in \bZ + \tilde{\nu} } \frac{ {\tilde{\rho}}^{A}_{r} }{{\bar{z}}^{r + \frac{1}{2}}}.
\end{eqnarray}
Here $\nu$ and $\tilde{\nu}$ labels the Ramond and Neveu-Schwarz sectors.  
The fields with $\nu=0$ and $\nu=\frac{1}{2}$ are the Ramond fields and Neveu-Schwarz fields, respectively.  
In a similar way, the fields with $\tilde{\nu}$ are defined.  
The commutators for these operators are 
\begin{eqnarray}
\{ \psi^{\mu}_r , \psi^{\nu}_s \} &=& \eta^{\mu \nu} \delta_{s+t,0}  \label{b1} ,\\
\{ {\tilde{\psi}}^{\mu}_r , {\tilde{\psi}}^{\nu}_s \} &=& \eta^{\mu \nu} \delta_{s+t,0} ,\\
\left[ \rho^A_r ,\rho^B_s \right] &=& i ~ G^{AB} \delta_{r+s,0} \label{b2} ,\\
\left[ {\tilde{\rho}}^A_r , {\tilde{\rho}}^B_s \right] &=& i ~ G^{AB} \delta_{r+s,0}. 
\end{eqnarray}
The energy-momentum tensor $T_B (z)$ and the worldsheet supercurrent $T_F (z)$ are 
\begin{eqnarray}
T_B (z) &=& - \frac{1}{{\alpha}'}  \partial X^{\mu} \partial  X_{\mu}  
     + i ~  \frac{1}{{\alpha}'} ~  \partial \Theta^{A} \partial  {\Theta}^{B} G_{BA} \nonumber \\
         & & - \frac{1}{2} \psi^\mu \partial \psi_\mu + \frac{i}{2} \rho^A \partial \rho_A  ,\\
T_F (z) &=& i \sqrt{\frac{2}{{\alpha}'}} \psi^\mu \partial X_\mu  + i \sqrt{\frac{2}{{\alpha}'}} \rho^A \partial \Theta_A .
\end{eqnarray}
As in the bosonic string case, the central charges for these matters can be computed and the results are $\frac{3}{2} (D_B - D_F)$.  
Therefore, after the superconformal ghosts are included, we can determine the critical dimensions as
\begin{equation}
D_B - D_F = 10  \label{122}.
\end{equation}
This implies that the critical dimensions are $(D_B,D_F)=(10,0),(12,2),(14,4),\cdots$.
Also, the Laurent expansions for $T_B$ and $T_F$ are
\begin{eqnarray}
T_B (z) &=& \sum_{m= - \infty}^{\infty} \frac{L_m}{z^{m+2}} ,\\
T_F (z) &=&  \sum_{r \in \bZ + \nu } \frac{G_r}{z^{r + \frac{3}{2}}} .
\end{eqnarray}
The super Virasoro operators are 
\begin{eqnarray}
L_n &=& \int \frac{d z}{2 \pi i } z^{n+1} T(z) \nonumber \\ 
    &=& \frac{1}{2} \sum_{m} \alpha^{\mu}_{n-m} \alpha_{\mu m}
      - \frac{i}{2} \sum_{m} \beta^{A}_{n-m} \beta^{B}_{ m} G_{BA} \nonumber \\
    & & +\frac{1}{4} \sum_{r \in \bZ + \nu }(2r -n ) \psi^{\mu}_{n-r} \psi_{\mu r} 
          -  \frac{i}{4} \sum_{r \in \bZ + \nu }(2r -n ) \rho^{A}_{n-r} \rho_{A r} ,\\
G_r &=& \sum_{m \in \bZ} \alpha^{\mu}_{m} \psi_{\mu r-m} + \sum_{m \in \bZ} \rho^A_{r-m} \beta_{m A} .
\end{eqnarray}
These operators satisfy the super Virasoro algebra.  
The zero-point energies for $L_0$ are $-\frac{1}{2}$ for NS and $0$ for R.

We discuss the physical states in the NS sectors and the R sectors.  
The ground states of open strings in the NS sectors are tachyons, 
and the massless states are gauge fields on supermanifolds.  
The interesting parts are the Ramond massless ground states, which are written as $| {\rm Ramond} ;p,w \rangle$ here.  
The Virasoro constraints are 
\begin{eqnarray}
0  &=& L_0 | {\rm Ramond} ;p,w \rangle =  \left( \frac{1}{2}  \alpha^{\mu}_{0} \alpha_{\mu 0}
      -  \frac{i}{2}  \beta^{A}_{0} \beta_{A 0}  \right) | {\rm Ramond} ;p,w \rangle ,\\
0 &=& G_0 | {\rm Ramond} ;p,w \rangle =  \left( \alpha^{\mu}_{0} \psi_{\mu 0} +  \rho^A_{0} \beta_{0 A} \right) | {\rm Ramond} ;p,w \rangle.
\end{eqnarray}
We may regard these equations as the Klein-Gordon equation and the Dirac equation on supermanifolds.  
The zero modes $\psi^{\mu}_0$ and $\rho^{A}_0$ of the Ramond sectors satisfy the following (anti)commutators;
\begin{eqnarray}
\{ \psi^{\mu}_0 , \psi^{\nu}_0 \} &=& \eta^{\mu \nu}   ,\\
\left[ \rho^A_0 ,\rho^B_0 \right] &=& i G^{AB} \label{123}.
\end{eqnarray}
The first anticommutators define the Clifford algebra and 
also the second commutators define the Heisenberg algebra.  
The spinor representation for the supergroup $Osp(1,D_B-1 | D_F)$ is an infinite dimensional representation because it includes the Heisenberg algebra.  
Thus, the Ramond massless ground states are labeled by this infinite dimensional spinor representation.  
Also, there are four sectors in the physical states of closed strings; NS-NS, R-R, R-NS and NS-R.

In order to extract the physical states, the GSO projection needs to be introduced.  
The GSO projection is defined by using the number operators defined below.  
For the fermionic fields $\psi^{\mu}$, the fermion number operator $F$ are defined as
\begin{equation}
(-1)^{F} \psi^{\mu} = - \psi^{\mu} (-1)^{F} .
\end{equation}
We can use this fermion number operator $F$ and the periodicity $\nu$ to extract the physical states for the parts of $\psi^{\mu}$.  
Here the tachyonic NS ground states are assigned as $(-1)^F=-1$.  
Next the number operator $F_{\rho}$ of $\rho^{A}$ is defined as
\begin{equation}
(-1)^{F_{\rho}} \rho^A = - \rho^A (-1)^{F_{\rho}} .
\end{equation}
All states can be labeled by these number operators .  For example the tachyonic ground states are represented by ${\rm NS}_{- +}$.  
The first label is the eigenvalue of $(-1)^{F}$ and the second label is the eigenvalue of $(-1)^{F_{\rho}}$.  
The massless states are written as ${\rm NS}_{+ +}$, ${\rm NS}_{- -}$, ${\rm R}_{+ \pm}$ and ${\rm R}_{- \pm}$.  
The number operators ${\tilde{F}}$ and ${{\tilde{F}}_{\tilde{\rho}}}$ for the right sectors are defined in the same way.

By using these operators, we can construct the type II string theories on supermanifolds as closed string theories.  
There are two kinds of GSO projections, which are called type IIA and type IIB.  
These projections satisfy the conditions that all pairs of vertex operators are mutually local, the operator product expansions are closed and 
the one loop amplitude is modular invariant.  Later we will show the modular invariance of the one loop amplitude explicitly.  
In type IIA string theory, the GSO projection is defined by keeping all sectors with
\begin{eqnarray}
\exp \left( i \pi  (F + F_{\rho}) \right) &=& 1 ,\\
\exp \left( i \pi  ({\tilde{F}} + {\tilde{F}}_{\rho}) \right) &=& (-1)^{1-2 {\tilde{\nu}}}.
\end{eqnarray}
In type IIB string theory, the GSO projection is defined by keeping all sectors with
\begin{eqnarray}
\exp \left( i \pi  (F + F_{\rho}) \right) &=& 1 ,\\
\exp \left( i \pi  ({\tilde{F}} + {\tilde{F}}_{\rho}) \right) &=& 1 .
\end{eqnarray}
The massless spectra of type II theory are divided by the four sectors NS-NS, R-R, NS-R and R-NS.  
The NS-NS massless states are a graviton, an antisymmetric two-form field and a dilaton on supermanifolds.  
The NS-R massless states have two space-time indices, which are the vector indices from the NS sector 
and the spinor indices from the R sector for the supergroup $Osp(1,D_B-1 |D_F)$.  
So the NS-R and R-NS massless states may be considered as two gravitini on supermanifolds.  
The R-R massless states have two spinor indices.  
If we use the matrix representations with two spinor indices for the generators of the Clifford algebra and the Heisenberg algebra,  
the R-R massless states are labeled by the tensors
\begin{eqnarray}
C_{\mu_1 \mu_2 \cdots \mu_{p+1} A_1 A_2 \cdots A_r } ,
\end{eqnarray}
which is antisymmetric for the indices $\mu_i$ of the bosonic coordinates and symmetric for the indices $A_s$ of the fermionic coordinates.  
These are equivalent to $(p+1|r)$-forms on supermanifolds.  
In \cite{a1}, the relation between the superpartners of the fermionic coordinates and the superforms is discussed in more detail by use of the picture changing operators.  
Also, the integers $p$ and $r$ are determined by the chiralities and the dimensions.  
The type IIA and the type IIB have different chiralities, and the chirality operator is defined by the product 
of the gamma matrix and the number operator of $\rho^A$.  
The number $p$ of bosonic coordinates are determined by the property of the spinor representation for the subgroup $SO(1,D_B-1)$ \cite{12}.  
Therefore, we can see that in the case that $\frac{D_B-2}{2}$ is even, in type IIA theory $(p,r)=({\rm even},{\rm even})$ and $(p,r)=({\rm odd},{\rm odd})$, 
and in type IIB theory $(p,r)=({\rm odd},{\rm even})$ and $(p,r)=({\rm even},{\rm odd})$.  
Moreover, in the case that $\frac{D_B-2}{2}$ is odd, in type IIA theory $(p,r)=({\rm odd},{\rm even})$ and $(p,r)=({\rm even},{\rm odd})$, 
and in type IIB theory $(p,r)=({\rm even},{\rm even})$ and $(p,r)=({\rm odd},{\rm odd})$.  
Later, we will use these R-R forms in order to construct D-branes on supermanifolds.

Also, we can construct the type 0A theory and the type 0B theory.  
The type 0A theory is defined by keeping all sectors with
\begin{eqnarray}
\nu &=& \tilde{\nu} ,\\
\exp \left( i \pi  (F + F_{\rho}) \right) &=& (-1)^{1-2 {\tilde{\nu}}} \exp \left( i \pi  ({\tilde{F}} + {\tilde{F}}_{\rho}) \right) .
\end{eqnarray}
Type 0B theory is defined by keeping all sectors with 
\begin{eqnarray}
\nu &=& \tilde{\nu} ,\\
\exp \left( i \pi  (F + F_{\rho}) \right) &=& \exp \left( i \pi  ({\tilde{F}} + {\tilde{F}}_{\rho}) \right) .
\end{eqnarray}
Moreover, the type I string theory can be defined by projecting type IIB theory by a worldsheet parity symmetry $\Omega$ 
and adding open string theory with gauge group $SO(32)$.

Finally,we discuss the modular invariance of the type II string theories on supermanifolds.  
The one-loop vacuum amplitude $Z_{T_2}$ is given by (\ref{1234}).  
\begin{equation}
Z_{T_2} = \int_{F_0} \frac{d^2 \tau}{4 \tau_2} Z_{X \Theta} (\tau) Z_{\psi \rho} (\tau) Z_{\tilde{\psi} \tilde{\rho}} (\bar{\tau}).
\end{equation}
The bosonic part $Z_{X \Theta}$ is 
\begin{equation}
Z_{X \Theta} = i  (4 \pi^2 {\alpha}'  )^{- \frac{D_B}{2} + \frac{D_F}{2}} ~ \frac{1}{\tau_2} 
             \left( \tau_2 (\eta(\tau) )^4 \right)^{- \frac{1}{2} (D_B - D_F -2)}.
\end{equation}
The part $Z_{\psi \rho} (\tau)$ of $\psi$ and $\rho$ is
\begin{eqnarray}
Z_{\psi \rho} (\tau) &=& \frac{1}{2} \left[ Z^0_0 (\tau)^{ \frac{D_B}{2} -1} {\tilde{Z}}^0_1 (\tau)^{\frac{D_F}{2}} - Z^0_1 (\tau)^{\frac{D_B}{2} -1} {\tilde{Z}}^0_0 (\tau)^{\frac{D_F}{2}} \right. \nonumber \\
              & &       \left.                     - Z^1_0 (\tau)^{ \frac{D_B}{2} -1} {\tilde{Z}}^1_1 (\tau)^{\frac{D_F}{2}} - Z^1_1 (\tau)^{ \frac{D_B}{2} -1} {\tilde{Z}}^1_0 (\tau)^{\frac{D_F}{2}}\right] . \label{c1}
\end{eqnarray}
Here $Z^{\alpha}_{\beta} (\tau)$ and ${\tilde{Z}}^{\alpha}_{\beta} (\tau) $ are defined by 
\begin{eqnarray}
Z^{\alpha}_{\beta} (\tau) &=& {\rm Tr }_{\alpha} \left( (-1)^{\beta F} q^{H_{\psi}} \right)  ,\\
{\tilde{Z}}^{\alpha}_{\beta} (\tau) &=& {\rm Tr }_{\alpha} \left( (-1)^{\beta F_{\rho}} q^{H_{\rho}} \right).
\end{eqnarray}
Here the label $\alpha$ is equal to $1- 2 \nu$.  $\alpha=1$ means in the R sector and $\alpha=0$ means in the NS sector.  
$H_{\psi}$ and $H_{\rho}$ are 
\begin{eqnarray}
H_{\psi} &=&  \frac{1}{4} \sum_{r \in \bZ + \nu } ( 2 r - n ) \psi^{j}_{n-r} \psi_{j r} + a_{\psi}  ,\\
H_{\rho} &=&  - \frac{i}{4} \sum_{r \in \bZ +\nu} (2r -n) \rho^a_{n-r} \rho_{a r} + a_{\rho}  ,
\end{eqnarray}
where $j=2 3,\cdots, D_B-1$ and $a = 1,2,\cdots, D_F$.  
The zero point energies $a_{\psi}$ and $a_{\rho}$ are 
\begin{eqnarray}
a_{\psi} &=& \frac{(1- 2\nu)^2}{8} - \frac{1}{24}  , \\
a_{\rho} &=& - \frac{(1- 2\nu)^2}{8} + \frac{1}{24} .
\end{eqnarray}
The part $Z^{\alpha}_{\beta} (\tau)$ of $\psi^{\mu}$ is completely the same as that of the usual type II superstring on the bosonic manifold. 
\begin{eqnarray}
Z^{\alpha}_{\beta} (\tau) &=& {\rm Tr }_{\alpha} \left( (-1)^{\beta F} q^{H_{\psi}} \right)  \nonumber \\
           &=& q^{\frac{ {\alpha}^2}{8}- \frac{1}{24}} \prod_{m=1}^{\infty} \left( 1 + (-1)^{\beta} q^{m-\frac{1-\alpha}{2}} \right) 
                               \left( 1 + (-1)^{\beta} q^{m- \frac{1 + \alpha}{2}} \right).
\end{eqnarray}
%
The part ${\tilde{Z}}^{\alpha}_{\beta} (\tau)$ of $\rho^A$ is
\begin{eqnarray}
{\tilde{Z}}^{\alpha}_{\beta} (\tau) &=& {\rm Tr }_{\alpha} \left( (-1)^{\beta F_{\rho}} q^{H_{\rho}} \right)   \nonumber \\
              &=& q^{- \frac{ {\alpha}^2}{8} + \frac{1}{24} } \prod_{m=1}^{\infty} \sum_{N_m, {\tilde{N}}_m =1 }^{\infty} \left( (-1)^{\beta} q^{m-\frac{1-\alpha}{2}} \right)^{N_m}  
                               \left( (-1)^{\beta} q^{m- \frac{1 + \alpha}{2}} \right)^{{\tilde{N}}_m}  \nonumber \\
&=& q^{- \frac{ {\alpha}^2}{8}  + \frac{1}{24} } \prod_{m=1}^{\infty} \left( 1 - (-1)^{\beta} q^{m-\frac{1-\alpha}{2}} \right)^{-1} 
                               \left( 1 - (-1)^{\beta} q^{m- \frac{1 + \alpha}{2}} \right)^{-1}  \nonumber \\
&=& \left( Z^{\alpha}_{\beta+1} (\tau) \right)^{-1} .
\end{eqnarray}
Here the parts of the ground states in the Ramond sectors are computed by use of the following expression; 
\begin{equation}
\left( \sum_{n=0}^{1} (\pm R)^n \right) \left( \sum_{n=0}^{\infty} (\mp R)^n \right) =1 , 
\end{equation}
where the parameter $R$ approaches to $1$.  
After all results are substituted into (\ref{c1}),
\begin{eqnarray}
Z_{\psi \rho} (\tau) &=& \frac{1}{2} \left[Z^0_0 (\tau)^{ \frac{D_B-2}{2}  } {\tilde{Z}}^0_1 (\tau)^{\frac{D_F}{2}} - Z^0_1 (\tau)^{\frac{D_B-2}{2}} {\tilde{Z}}^0_0 (\tau)^{\frac{D_F}{2}} \right. \nonumber \\
              & &       \left.                     - Z^1_0 (\tau)^{ \frac{D_B-2}{2}} {\tilde{Z}}^1_1 (\tau)^{\frac{D_F}{2}} - Z^1_1 (\tau)^{ \frac{D_B-2}{2} } {\tilde{Z}}^1_0 (\tau)^{\frac{D_F}{2}} \right] \nonumber \\
&=& \frac{1}{2} \left[  Z^0_0 (\tau)^{ \frac{D_B - D_F-2}{2}} - Z^0_1 (\tau)^{\frac{D_B - D_F-2}{2}} \right. \nonumber \\
              & &       \left.    - Z^1_0 (\tau)^{ \frac{D_B - D_F-2}{2}} - Z^1_1 (\tau)^{ \frac{D_B -D_F-2}{2}} \right].
\end{eqnarray}
Similarly we can compute the part $Z_{\tilde{\psi} \tilde{\rho}} (\bar{\tau})$.  
\begin{eqnarray}
Z_{\tilde{\psi} \tilde{\rho}} (\bar{\tau}) &=& 
\frac{1}{2} \left[  Z^0_0 (\bar{\tau})^{ \frac{D_B - D_F-2}{2}} - Z^0_1 (\bar{\tau})^{\frac{D_B - D_F-2}{2}} \right. \nonumber \\
              & &       \left.    - Z^1_0 (\bar{\tau})^{ \frac{D_B - D_F-2}{2}} \mp Z^1_1 (\bar{\tau})^{ \frac{D_B -D_F-2}{2}} \right] .
\end{eqnarray}
For the signs of the last term, the negative sign corresponds to the type IIB theory and the positive sign corresponds to the type IIA theory.  
Therefore, we can see that the modular parameter dependent parts of these one-loop vacuum amplitudes 
are completely the same as those of the usual type II superstring theories on the flat bosonic manifold if the condition (\ref{122}) is satisfied.  
In conclusion, the type II string theories on supermanifolds with vanishing central charges are modular invariant.

\subsection{\Large Heterotic String Theories on Flat Supermanifolds}

We study heterotic string theories on flat supermanifolds in the same way as $SO(32)$ and $E_8 \times E_8$ 
heterotic string theories on the flat bosonic manifold.  The heterotic string theories are constructed by combining 
the bosonic string theories as the left-moving part and the type II string theories as the right-moving part.  
As discussed before, we have constructed many bosonic string theories and type II string theories on different supermanifolds.  
So we can construct many different heterotic string theories.  We consider the heterotic string theories 
whose left-handed parts are the bosonic string theories with the dimension $(D_B^L,D_F^L)$ and 
whose right-handed parts are the RNS string theories with the dimension $(D_B^R,D_F^R)$.  
In the case that $D_F^L$ is larger than $D_F^R$, the actions for the heterotic string theories on flat supermanifolds are 
\begin{eqnarray}
S &=& \frac{1}{4 \pi} \int d^2 z  \left( \frac{2}{{\alpha}'}  \partial X^{\mu} \bar{\partial} X_{\mu} 
- i \frac{2}{{\alpha}'}  \partial \Theta^A \bar{\partial} \Theta_A \right.\nonumber \\
    & & \left. ~~~~~~ +  \lambda^{i} \bar{\partial} \lambda^i +  \lambda^{i'} \bar{\partial} \lambda^{i'} +  {\tilde{\psi}}^{\mu} \partial {\tilde{\psi}}_\mu 
             - i  \pi^a \bar{\partial} \pi_a  - i  {\tilde{\rho}}^A \partial {\tilde{\rho}}_A \right),
\end{eqnarray}
where $\lambda^{i}$ and $\lambda^{i'}$ are Grassmann odd fields for $i=17,18,\cdots,2 D_B^L - 20$ and $i'=1,2,\cdots,16$, 
and $\pi^a$ are Grassmann even fields for $a=1,2,\cdots,2 D_F^L$.  
The operator product expansions and the energy-momentum tensors for these fields are 
\begin{eqnarray}
& & \lambda^i (z) \lambda^j (0) \sim \frac{\delta^{ij}}{z} , ~~~~ \lambda^{i'} (z) \lambda^{j'} (0) \sim \frac{\delta^{i'j'}}{z} , \\
& & \pi^a (z) \pi^b (0) \sim \frac{ i G^{ab}}{z}  ,\\
& & T_{\lambda} (z) = - \frac{1}{2} \lambda^i  \partial \lambda^i , ~~~~ T_{{\lambda}'} (z) = - \frac{1}{2} \lambda^{i'}  \partial \lambda^{i'} \\
& & T_{\pi} (z) =   \frac{i}{2} \pi^a \partial \pi_a  .
\end{eqnarray}
The left-dimension $D_B^L | D_F^L$ and the right-dimension $D_B^R | D_F^R$ should be chosen so that each central charge is zero. 
\begin{eqnarray}
D_B^L - D_F^L &=& 26 \nonumber ,\\
D_B^R - D_F^R &=& 10.
\end{eqnarray}
Here we concretely consider only the simplest cases that $(D_B^R, D_F^R ) = (10,0)$.  
In these cases, the right-moving massless states are labeled by the vector representation $8_v$ and the spinor representation $8_s$ for $SO(8)$, 
where the massless physical states are classified by their behavior under the SO(8) rotations as usual \cite{12}.  
The number operators $F_{\lambda}$, $F_{{\lambda}'}$ and $F_{\pi}$ for the left-moving fields are introduced 
in order to define the GSO projection on $\lambda^i$, $\lambda^{i'}$ and $\pi^a$.  
%
\begin{eqnarray}
(-1)^{F_{\lambda}} \lambda^i &=& - \lambda^i (-1)^{F_{\lambda}} ,\\
(-1)^{F_{{\lambda}'}} \lambda^{i'}&=& - \lambda^{i'} (-1)^{F_{{\lambda}'}} ,\\
(-1)^{F_{\pi}} \pi^a &=& - \pi^a (-1)^{F_{\pi}}.
\end{eqnarray}
We take the GSO projection 
\begin{equation}
\exp \left( i \pi (F_{\lambda} + F_{{\lambda}'} + F_{\pi} ) \right) =1.
\end{equation}
This GSO projection satisfies the consistency conditions such as mutual locality and modular invariance.  
The left-moving massless spectra of this theory are the states labeled by the vector representation $8_v$ 
and the states for the adjoint representation of $Osp(2 D^L_B-20| 2 D^L_F)$:  
\begin{eqnarray}
\alpha_{-1}^{\mu} | 0 \rangle_{{\rm NS}} ,~~& & ~~~ \lambda^i_{-\frac{1}{2}} \lambda^j_{-\frac{1}{2}} | 0 \rangle_{{\rm NS}} \nonumber \\
\lambda^i_{- \frac{1}{2}} \pi^b_{- \frac{1}{2}} | 0 \rangle_{{\rm NS}}, ~~ & & \pi^a_{- \frac{1}{2}} \lambda^j_{- \frac{1}{2}} | 0 \rangle_{{\rm NS}} 
, ~~~~~ \pi^a_{- \frac{1}{2}} \pi^b_{- \frac{1}{2}} | 0 \rangle_{{\rm NS}} .
\end{eqnarray}
The parts with the label $i'$ are the same as those with the label $i$.  
Therefore by combining right-moving and left-moving parts, we can see that the massless states are related to the $N=1$ supergravity 
and the $N=1$ gauge multiplets with the gauge group $Osp(2 D^L_B-20| 2 D^L_F)$ on the 10 dimensional flat bosonic Minkowski space.

Moreover, other heterotic string theories on supermanifolds can be constructed.  
For example, the GSO projection 
\begin{eqnarray}
& & \exp \left( i \pi (F_{\lambda} + F_{\pi} ) \right) =1,  \nonumber \\
& & \exp \left( i \pi (F_{{\lambda}'} ) \right) =1
\end{eqnarray}
is taken to satisfy the consistency conditions similarly.  
There are four sectors in the physical spectra, because $(\lambda^i,\pi^a)$ and $\lambda^{i'}$ can have independent boundary conditions, respectively;
\begin{eqnarray}
\lambda^i (w + 2 \pi ) = \eta ~ \lambda^i (w) \nonumber , \\
\pi^a  (w + 2 \pi ) = {\eta}  ~ \pi^a (w) \nonumber ,\\
\lambda^{i'} (w + 2 \pi ) = \tilde{\eta}  ~ \lambda^{i'} (w) .
\end{eqnarray}
Here we use the worldsheet coordinate $w=\sigma^1 + i \sigma^2$, and the periodicity $\eta=\pm 1$ and $\tilde{\eta}=\pm 1$.  
These four sectors are labeled by NS-NS',  R-NS', NS-R' and R-R'.  
The left-moving massless spectra of this theory are the NS-NS' parts 
\begin{eqnarray}
\alpha_{-1}^{\mu} | 0 \rangle_{{\rm NS}-{\rm NS}'} ,~~& & \lambda^{i'}_{-\frac{1}{2}} \lambda^{j'}_{-\frac{1}{2}} | 0 \rangle_{{\rm NS}-{\rm NS}'}, ~~~
\lambda^i_{-\frac{1}{2}} \lambda^j_{-\frac{1}{2}} | 0 \rangle_{{\rm NS}-{\rm NS}'} , \nonumber \\
\lambda^i_{- \frac{1}{2}} \pi^b_{- \frac{1}{2}} | 0 \rangle_{{\rm NS}-{\rm NS}'}, ~~ & & \pi^a_{- \frac{1}{2}} \lambda^j_{- \frac{1}{2}} | 0 \rangle_{{\rm NS}-{\rm NS}'} 
, ~~~ \pi^a_{- \frac{1}{2}} \pi^b_{- \frac{1}{2}} | 0 \rangle_{{\rm NS}-{\rm NS}'} ,
\end{eqnarray}
and the R-NS' parts and the NS-R' parts labeled by 
\begin{eqnarray}
& & \{ \lambda^{i}_{0}, \lambda^{j}_{0} \} = \delta^{i j} ,~~~~ \left[ \pi^a_0 , \pi^b_0 \right] = i G^{ab} ,\nonumber \\
& &  \{ \lambda^{i'}_{0}, \lambda^{j'}_{0} \} = \delta^{i' j'} .
\end{eqnarray}
Therefore, the massless spectra are labeled by the sums of the adjoint representations and the spinor representations of $Osp(2 D_B^L - 36 | 2 D_F^L )$ and $SO(16)$.  
In order to see if these massless states are the $N=1$ gauge multiplets on the 10 dimensional flat bosonic Minkowski space, 
it would also be interesting to check that 
the sums of the adjoint representations and the spinor representations of $Osp(2 D_B^L - 36 | 2 D_F^L )$ are 
infinite dimensional supergroups.

\section{\Large D-branes on Supermanifolds}

In this section, we study D-branes on supermanifolds as non-perturbative objects in string theory.  
In the case of supermanifolds, we can impose the boundary conditions for the fermionic coordinates.  
So, D-branes on supermanifolds can have more configurations than those on bosonic manifolds.  
Boundary states are constructed to compute the amplitudes in the closed string channel.  
Here we obey the notations in the papers \cite{9}.  

\subsection{\Large D-branes in Bosonic String Theories on Supermanifolds}

We construct D-branes in the bosonic string theory on supermanifolds.  The dimensions of the supermanifolds are written as $D_B|D_F$.  
The critical dimensions are $D_B -D_F=26$, that is, $(D_B,D_F) =(26,0),(28,2),(30,4), \cdots$.  
In order to construct D-branes, we study the boundary conditions on the worldsheet.  
The Dirichlet condition or the Neumann condition for each bosonic direction is imposed on the boundary.  
But we have to treat the fermionic parts $\Theta^{A}$ carefully, because the two fields $\Theta^{A}$ and $\Theta^{A+1}$ appear in pairs in the action.  
The same boundary conditions for each pair of fermionic directions need to be imposed, because the following conditions should be satisfied on the boundary;
\begin{eqnarray}
& & \delta \Theta^A ~ \partial_{\sigma^1} \Theta^{A+1} = 0 ,\nonumber \\
& & \delta \Theta^{A+1} ~ \partial_{\sigma^1} \Theta^{A} = 0 . \label{3000}
\end{eqnarray}
Therefore we can see that from these boundary conditions, $\delta \Theta^A = \delta \Theta^{A+1}=0$ or $\partial_{\sigma^1} \Theta^{A} = \partial_{\sigma^1} \Theta^{A+1}= 0$.  
When we denote the numbers of the directions with the Neumann conditions as $p+1$ for the bosonic directions and $r$ for the fermionic directions, 
$r$ has to be an even number.  
We call these D-branes as ${\rm D}_{p|r}$-branes.

We compute the amplitude $F$ between two parallel ${\rm D}_{p|r}$-branes in the open string channel and in the closed string channel to determine the tensions of the ${\rm D}_{p|r}$-branes.  
At first in the open string channel, the one loop open string amplitude is
\begin{equation}
F = \int^{\infty}_{0} \frac{d \tau}{2 \tau} {\rm Tr } ~ e^{- 2 \pi L_0 \tau}  .
\end{equation}
Here we denote the length of the cylinder as $\tau$.  The trace is taken over the open string physical states and noncompact momenta.  
The mode expansions of the fields $X^{\alpha}$ and $\Theta^a$ for the NN directions are 
\begin{eqnarray}
X^{\alpha} &=& x^{\mu} -i {\alpha}' p^{\mu} \log |z|^2 + i \left( \frac{{\alpha}'}{2} \right)^{\frac{1}{2} } \sum_{m \not= 0 } \frac{{\alpha^\mu_m}}{m} 
\left( z^{-m} + {\bar{z}}^{-m} \right)  ,\\
\Theta^a &=& \theta^a - i {\alpha}' w^a \log |z|^2 + i \left( \frac{{\alpha}'}{2} \right)^{\frac{1}{2} } \sum_{m \not= 0 } \frac{\beta^a_{m}}{m} 
\left( z^{-m} + {\bar{z}}^{-m} \right)  .
\end{eqnarray}
The mode expansions of the fields $X^{i}$ and $\Theta^s$ for the DD directions are 
\begin{eqnarray}
X^i &=&  - i \frac{y^i}{2 \pi}  \log \left(\frac{z}{\bar{z}} \right) + i \sqrt{\frac{{\alpha}'}{2}} \sum_{m \neq 0} \frac{\alpha^i_m}{m} \left( z^{-m} - {\bar{z}}^{-m} \right) ,\\
\Theta^s &=& i \sqrt{\frac{{\alpha}'}{2}} \sum_{m \neq 0} \frac{\beta^s_m}{m} \left( z^{-m} - {\bar{z}}^{-m} \right),
\end{eqnarray}
where $y^i$ is a distance between two parallel ${\rm D}_{{ p|r}}$-branes.  
In the DD fermionic directions, the ${\rm D}_{{ p|r}}$-branes are localized at $\theta^s=0$.  
The Virasoro generator $L_0$ is 
\begin{equation}
L_0 = {\alpha}' (k^{\alpha} k_{\alpha} - i w^a w_a ) + \frac{y^2}{4 \pi^2 {\alpha}'} + \sum_{n=1}^{\infty} ( \alpha^{\mu}_{-n} \alpha_{\mu n} - i \beta^A_{-n} \beta_{An} )   
+ L_0^{\rm ghost} .
\end{equation}
The amplitude $F$ can be computed as
\begin{eqnarray}
F &=& 2 i \int \frac{ d^{p+1} k}{(2 \pi)^{p+1}} (-i 2 \pi)^{\frac{r}{2}} \int d^r w \int^{\infty}_{0} \frac{d \tau}{2 \tau} 
e^{ - 2 \pi \tau {\alpha}' (k^2 - i w^A w_A )} ~ e^{- \frac{y^2 \tau}{2 \pi {\alpha}'}} e^{2 \pi \tau} 
~ {\rm Tr} \left( \prod^{\infty}_{n=1} e^{N^{\perp}} \right) \nonumber \\
&=& i  \int^{\infty}_{0} d \tau ( 8 \pi^2 {\alpha}' \tau )^{- \frac{p+1}{2} + \frac{r}{2}} 
~ e^{- \frac{y^2 \tau}{2 \pi {\alpha}'}} e^{2 \pi \tau} 
\prod^{\infty}_{n=1} \left( \frac{1}{1- e^{-2 \pi \tau n} } \right)^{D_B - D_F -2}  \nonumber \\
&=& i  ( 8 \pi^2 {\alpha}'  )^{- \frac{p+1}{2} + \frac{r}{2}} 
 \int^{\infty}_{0} \frac{d \tau}{\tau} {\tau}^{- \frac{p+1}{2}+\frac{r}{2}} ~ e^{- \frac{y^2 \tau}{2 \pi {\alpha}'}}
\left( f_1 (e^{ -\pi \tau}) \right)^{- 24} \label{01}.
\end{eqnarray}
Here $N^{\perp}$ is defined by
\begin{equation}
N^{\perp} = \sum_{n=1}^{\infty} ( \alpha^{j}_{-n} \alpha_{j n} - i \beta^A_{-n} \beta_{An} )  , 
\end{equation}
where $j$ runs over $2,3,\cdots,D_B-1$.  The function $f_1$ is defined as
\begin{equation}
f_1 (q) = q^{\frac{1}{12}} \prod^{\infty}_{n=1} (1 - q^{2n}) . 
\end{equation}
%
%
Later, the amplitude (\ref{01}) in the open string channel will be compared with the amplitude in the closed string channel.

Next the amplitude of the same configuration is computed in the closed string channel.  
We construct the boundary states as
\begin{equation}
|B \rangle = \frac{T_{p,r}}{2} |B_X \rangle |B_{\Theta} \rangle  |B_{\rm ghost} \rangle ,
\end{equation}
where $T_{p,r}$ is a normalization constant to be fixed.  The boundary states for the bosonic coordinates are defined by the following conditions;
\begin{eqnarray}
& & \partial_{\sigma_2} X^{\alpha} |_{\sigma_2=0} |B_X \rangle =0 ~~~~~~~~~~~~~~  \alpha = 0,\cdots,p  ,\\
& & X^i |_{\sigma_2 = 0} |B_X \rangle = y^i |_{\sigma_2=0} |B_X \rangle ~~~~~~ i = p+1, \cdots , D_B-1    .
\end{eqnarray}
These conditions can be written by use of the oscillators;
\begin{eqnarray}
& & (\alpha_n^{\alpha} + {\tilde{\alpha}}^{\alpha}_{-n} ) |B_X \rangle =0 ,~~~
(\alpha_n^{i} - {\tilde{\alpha}}^{i}_{-n} )  |B_X \rangle =0 ,\label{02}\\
& & {\hat{p}}^{\alpha} |B_X \rangle =0 , ~~~~~~({\hat{x}}^{i} -  y^i ) |B_X \rangle =0 .
\end{eqnarray}
Here a matrix $S^{\mu \nu}$ is defined by
\begin{equation}
S^{\mu \nu} = ( \eta^{\alpha \beta} , - \delta^{ij}) .
\end{equation}
The conditions (\ref{02}) are rewritten as
\begin{equation}
(\alpha_n^{\mu} + {S^{\mu}}_{ \nu} {\tilde{\alpha}}^{\nu}_{-n} ) |B_X \rangle =0 .
\end{equation}
The boundary states satisfying these conditions can be constructed as 
\begin{equation}
|B_X \rangle = \delta^{D_B -p-1} ( {\hat{x}}^{i} -  y^i ) \left( \prod_{n=1}^{\infty} 
e^{- \frac{1}{n} \alpha_{-n}^{\mu} S_{\mu \nu} {\tilde{\alpha}}_{-n}^{\mu}} \right)
|0 \rangle_{\alpha} |0 \rangle_{\tilde{\alpha}} |p^{\alpha} =0 \rangle .
\end{equation}
Here $|0 \rangle_{\alpha}$ and $ |0 \rangle_{\tilde{\alpha}}$ are defined by 
\begin{eqnarray}
& & \alpha^{\mu}_{n} |0 \rangle_{\alpha} = 0  ~~~~ {\rm for} ~~ n >0 , \nonumber \\
& & {\tilde{\alpha}}^{\mu}_{n} |0 \rangle_{\tilde{\alpha}} = 0 ~~~~ {\rm for} ~~ n >0 .
\end{eqnarray}
Next we construct the boundary states for the fermionic coordinates by imposing the following conditions;
\begin{eqnarray}
& & \partial_{\sigma_2} {\Theta}^{a} |_{\sigma_2=0} |B_{\Theta} \rangle =0 ~~~~~~ a = 1,\cdots, r  ,\\
& & {\Theta}^s |_{\sigma_2= 0} |B_{\Theta} \rangle = 0 ~~~~~~ s = r+1, \cdots , D_F  .
\end{eqnarray}
These conditions can be written by use of the oscillators;

\begin{eqnarray}
& & (\beta_n^{a} + {\tilde{\beta}}^{a}_{-n} ) |B_{\Theta} \rangle =0 ,~~~
(\beta_n^{s} - {\tilde{\beta}}^{s}_{-n} )  |B_{\Theta} \rangle =0 \label{03} ,\\
& & {\hat{w}}^{a} |B_{\Theta} \rangle =0 , ~~~~~~  {\hat{\theta}}^{s}   |B_{\Theta} \rangle =0 .
\end{eqnarray}
We define $S^{A B} = (  G^{a b } , - G^{s t}) $, and then the conditions (\ref{03}) are rewritten as 
\begin{equation}
(\beta_n^{A} + S^{A B} {\tilde{\beta}}_{B -n} ) |B_{\Theta} \rangle =0.
\end{equation}
The boundary states satisfying these conditions can be constructed as 
\begin{equation}
|B_{\Theta} \rangle = \delta^{D_F-r} ( {\hat{\theta}}^{s} )  \prod_{n=1}^{\infty} \left( e^{ \frac{i}{n} \beta_{-n}^{A} 
{S_{A}}^{ B} {\tilde{\beta}}_{B -n}} \right)
|0 \rangle_{\beta} |0 \rangle_{\tilde{\beta}} |w^a =0 \rangle ,
\end{equation}
where $|0 \rangle_{\beta}$ and $ |0 \rangle_{\tilde{\beta}}$ are defined by 
\begin{eqnarray}
& & \beta^A_{n} |0 \rangle_{\beta} = 0 ~~~~ {\rm for} ~~ n > 0 \nonumber , \\
& & {\tilde{\beta}}^A_{n} |0 \rangle_{\tilde{\beta}} = 0 ~~~~ {\rm for} ~~ n > 0 .
\end{eqnarray}
The ghost part is the same as usual,
\begin{equation}
|B_{\rm ghost} \rangle = e^{\sum^{\infty}_{n=0} (c_{-n} {\tilde{b}}_{-n} - b_{-n} {\tilde{c}}_{-n} )} \left( \frac{c_0 + {\tilde{c}}_0 }{2} \right) 
| q=1 \rangle | \tilde{q} =1 \rangle ,
\end{equation}
where the state $| q=1 \rangle$ is defined as
\begin{eqnarray}
& & c_n | q=1 \rangle =0 ~~~ {\rm for} ~~~ n \ge 1  , \nonumber \\
& & b_m | q=1 \rangle =0 ~~~  {\rm for} ~~~ m \ge 0 ,
\end{eqnarray}
and the antiholomorphic part $| \tilde{q} =1 \rangle$ is defined in the same way.  
%
%

The amplitude $F$ between two parallel ${\rm D}_{p|r}$-branes in the closed string channel can be computed by use of these boundary states.  
\begin{equation}
F = \langle B | D | B \rangle ,
\end{equation}
where the closed string propagator is defined as
\begin{equation}
D = \frac{{\alpha}'}{4 \pi}  \int_{1 \ge |z|} \frac{d^2 z}{|z|^2} ~ z^{L_0} {\bar{z}}^{{\tilde{L}}_0} .
\end{equation}
The amplitude is factorized as follows;
\begin{eqnarray}
F = \frac{{\alpha}'}{4 \pi} \frac{T^2_{p,r}}{4}        \int_{1 \ge |z|} \frac{d^2 z}{|z|^2} Z_{X}^{{\rm zero}} Z_{X}^{{\rm nonzero}}
Z_{\Theta}^{{\rm zero}} Z_{\Theta}^{{\rm nonzero}} Z_{{\rm Ghost} } \label{eq1}.
\end{eqnarray}
The zero mode part of $X^{\mu} $ is 
\begin{eqnarray}
Z_{X}^{{\rm zero}} &=& \langle p=0 | \delta^{d_{\perp} } ( \hat{q_i }) |z|^{\frac{{\alpha}'}{2} {\hat{p}}^2 }\delta^{d_{\perp}} ( {\hat{q}}_i - y_i ) | p = 0 \rangle  \nonumber \\
&=&  i ( 2 \pi^2 {\alpha}' t )^{- \frac{d_{\perp} }{2}} e^{ - \frac{y^2}{2\pi {\alpha}' t}}  .
\end{eqnarray}
Here we define $d_{\perp} = D_B - 1 - p$, and the parameter $t$ is defined by 
\begin{equation}
|z| = e^{- \pi t} ~~, ~~ d^2 z = - \pi e^{ - 2 \pi t } d t d \phi .
\end{equation}
The nonzero mode part of $X^{\mu}$ is 
\begin{eqnarray}
Z_{X}^{{\rm nonzero}} &=&  {}_{\alpha} \langle 0| {}_{\tilde{\alpha}} \langle 0| 
\left( \prod_{m=1}^{\infty} 
e^{- \frac{1}{m} \alpha_{-m}^{\mu} S_{\mu \nu} {\tilde{\alpha}}_{-m}^{\mu}} \right)
z^N {\bar{z}}^{\tilde{N}} 
\left( \prod_{n=1}^{\infty} 
e^{- \frac{1}{n} \alpha_{-n}^{\mu} S_{\mu \nu} {\tilde{\alpha}}_{-n}^{\mu}} \right)
|0 \rangle_{\alpha} |0 \rangle_{\tilde{\alpha}}  \nonumber \\
&=&  \prod^{\infty}_{n=1} \left( \frac{1}{1- |z|^{2n}} \right)^{D_B}  ,
\end{eqnarray}
where the $N$ and $\tilde{N}$ are the number operators.
\begin{eqnarray}
N &=& \sum^{\infty}_{n=1} \alpha^{\mu}_{-n} \alpha_{\mu n} - i \sum^{\infty}_{m=1} \beta_{-m}^{A} \beta_{A m}         ,\\
\tilde{N} &=&  \sum^{\infty}_{n=1} {\tilde{\alpha}}^{\mu}_{-n} {\tilde{\alpha}}_{ \mu n} 
- i \sum^{\infty}_{m=1} {\tilde{\beta}}_{-m}^{A} {\tilde{\beta}}_{A m} .
\end{eqnarray}
The zero mode part of $\Theta^A$ is 
\begin{eqnarray}
Z_{\Theta}^{{\rm zero}} &=& \langle w=0 | \delta ( \hat{\theta} ) ~ |z|^{- i \frac{{\alpha}'}{2} w^A w_A } ~ 
\delta ( \hat{\theta} ) | w=0 \rangle \nonumber \\
&=& (2 \pi^2 {\alpha}' t )^{\frac{D_F - r}{2}}.
\end{eqnarray}
The nonzero mode part of $\Theta^A$ is 
\begin{eqnarray}
Z_{\Theta}^{{\rm nonzero}} &=& {}_{\beta} \langle 0| {}_{\tilde{\beta}} \langle 0| 
\prod_{n=1}^{\infty} \left( e^{ \frac{i}{n} \beta_{n}^{A} 
{S_{A}}^{ B} {\tilde{\beta}}_{B n}} \right)
z^N {\bar{z}}^{\tilde{N}} 
\prod_{n=1}^{\infty} \left( e^{ \frac{i}{n} \beta_{-n}^{A} 
{S_{A}}^{ B} {\tilde{\beta}}_{B -n}} \right)
|0 \rangle_{\beta} |0 \rangle_{\tilde{\beta}} \nonumber \\
&=&  \prod^{\infty}_{n=1} \left( 1- |z|^{2n} \right)^{D_F}  .
\end{eqnarray}
The ghost part is 
\begin{eqnarray}
Z_{{\rm Ghost} } = |z|^{-2}  \prod^{\infty}_{n=1} \left( 1- |z|^{2n} \right)^{2} .
\end{eqnarray}
After these results are substituted into (\ref{eq1}), the amplitude is
\begin{eqnarray}
F &=& \frac{T^2_{p,r}}{4} \frac{i {\alpha}' \pi}{2} \int^{\infty}_{0} dt ~ e^{2 \pi t} ~ ( 2 \pi^2 {\alpha}' t )^{- \frac{d_{\perp} }{2}  + \frac{D_F -r}{2}} 
~ e^{ - \frac{y^2}{2\pi {\alpha}' t}} 
\prod^{\infty}_{n=1} \left( \frac{1}{1- e^{-2 \pi n t}} \right)^{D_B-D_F-2} \nonumber \\
&=& \frac{T^2_{p,r}}{4} \frac{i {\alpha}' \pi}{2} ( 2 \pi^2 {\alpha}' )^{- \frac{d_{\perp} }{2} + \frac{D_F -r}{2}} 
\int^{\infty}_{0} dt ~ 
t^{- \frac{d_{\perp}}{2}  + \frac{D_F -r}{2}} ~ e^{ - \frac{y^2}{2\pi {\alpha}' t}}  
\left( f_1 (e^{ -\pi t}) \right)^{- 24} \label{04}.
\end{eqnarray}

Finally we compare the amplitude (\ref{01}) in the open string channel and the amplitude (\ref{04}) in the closed string channel.  The well-known identity for the function $f_1$ is
\begin{eqnarray}
\tau &=& \frac{1}{t} ,\\
f_1 (e^{ - \pi \tau })  &=&  f_1 (e^{ - \frac{\pi}{t}  }) =  \sqrt{t} f_1 (e^{ -\pi t}).
\end{eqnarray}
After using this identity, the amplitude (\ref{01}) is rewritten by use of the parameter $t$ instead of $\tau$.
\begin{equation}
F= i  ( 8 \pi^2 {\alpha}'  )^{- \frac{p+1}{2}+ \frac{r}{2}} 
\int^{\infty}_{0} dt ~ 
t^{- \frac{d_{\perp} }{2} + \frac{D_F -r}{2}} e^{ - \frac{y^2}{2\pi {\alpha}' t}}  
\left( f_1 (e^{ -\pi t}) \right)^{- 24}.
\end{equation}
In conclusion, we can see that the amplitudes in the open and closed string channels are completely the same if the tension of ${\rm D}_{p|r}$-brane is defined as
\begin{equation}
T_{p,r} = \sqrt{\pi} ~ 2^{- \frac{D_B - D_F - 10}{4}} ~ 
(2 \pi \sqrt{{\alpha}'} )^{\frac{D_B}{2} - \frac{D_F}{2} -2 - p   + r }  .
\end{equation}
The NN fermionic directions contribute to the tension as an inverse power of the NN bosonic directions.  
%
~ \\
~ \\
~ \\
\newpage

\subsection{\Large D-branes in RNS String Theories on Supermanifolds}

Here we consider D-branes in the type II string theories on supermanifolds.  
We remember two facts.  One is that the R-R $(p+1,r)$-forms exist in the type II string theories on supermanifolds.  
In the case that $\frac{D_B-2}{2}$ is even, in type IIA theory $(p,r)=({\rm even},{\rm even})$ and $(p,r)=({\rm odd},{\rm odd})$, 
and in type IIB theory $(p,r)=({\rm odd},{\rm even})$ and $(p,r)=({\rm even},{\rm odd})$.  
Also, in the case that $\frac{D_B-2}{2}$ is odd, in type IIA theory $(p,r)=({\rm odd},{\rm even})$ and $(p,r)=({\rm even},{\rm odd})$, 
and in type IIB theory $(p,r)=({\rm even},{\rm even})$ and $(p,r)=({\rm odd},{\rm odd})$.  
Moreover, the other fact is that $r$ has to be an even number from (\ref{3000}).  
From these facts, we can see that in the case that $\frac{D_B-2}{2}$ is even, in type IIA theory the R-R forms with $(p,r)=({\rm even},{\rm even})$ 
and in type IIB theory the R-R forms with $(p,r)=({\rm odd},{\rm even})$ can couple to ${\rm D}_{p|r}$-branes.  
Also, in the case that $\frac{D_B-2}{2}$ is odd, in type IIA theory the R-R forms with $(p,r)=({\rm odd},{\rm even})$ 
and in type IIB theory the R-R forms with $(p,r)=({\rm even},{\rm even})$ can couple to ${\rm D}_{p|r}$-branes.  
Here $r=0,2,4,\cdots, D_F$.  
Now we will check what kinds of configurations satisfy the BPS-like conditions.  Here we limit the cases that D-branes are static and rigid, 
although we can consider the cases that D-branes are rotated and boosted \cite{9} \cite{a1}.  
We write the Dirichlet-Dirichlet directions as DD, Neumann-Neumann as NN, Neumann-Dirichlet as ND and Dirichlet-Neumann as DN, for short.

In the open string channel, we compute the amplitude between one ${\rm D}_{p|r}$-brane and one ${\rm D}_{q|\tilde{r}}$-brane.  
The numbers of the DD directions are denoted by $DD_B$ for the bosonic coordinates and $DD_F$ for the fermionic coordinates.  
The numbers of the NN directions are denoted by $NN_B$ and $NN_F$.  
The numbers of the ND and DN directions are denoted by $\nu_B$ and $\nu_F$.  
The boundary conditions for the ND directions are given by 
\begin{eqnarray}
\partial_{\sigma^1} X^\mu |_{\sigma^1 =0} = 0 , ~~ & & ~~ \partial_{\sigma^2} X^{\mu} |_{\sigma^1=\pi} = 0  ,\\
\partial_{\sigma^1} {\Theta}^A |_{\sigma^1 =0} = 0 , ~~ & & ~~ \partial_{\sigma^2} {\Theta}^{A} |_{\sigma^1=\pi} = 0  ,\\
\psi^{\mu} |_{\sigma^1 =0} = {\tilde{\psi}}^{\mu} |_{\sigma^1 =0} ,  ~~ & & ~~ \psi^{\mu} |_{\sigma^1 =\pi} = - e^{i 2 \pi \nu} {\tilde{\psi}}^{\mu} |_{\sigma^1 =\pi},  \\
\rho^{A} |_{\sigma^1 =0} = {\tilde{\rho}}^{A} |_{\sigma^1 =0} , ~~ & & ~~ \rho^{A} |_{\sigma^1 =\pi} = - e^{i 2 \pi \nu} {\tilde{\rho}}^{A} |_{\sigma^1 =\pi}.
\end{eqnarray}
The mode expansions for the ND directions are 
\begin{eqnarray}
X^{\mu} &=& i \left( \frac{{\alpha}'}{2} \right)^{\frac{1}{2}} \sum_{r \in \bZ + \frac{1}{2} } \frac{\alpha^{\mu}_{r}}{r} \left( z^{-r} + {\bar{z}}^{-r} \right) , \\
{\Theta}^A &=& i \left( \frac{{\alpha}'}{2} \right)^{\frac{1}{2}} \sum_{r \in \bZ + \frac{1}{2} } \frac{\beta^{A}_{r}}{r} \left( z^{-r} + {\bar{z}}^{-r} \right) , \\
\psi^{\mu} ( z) &=& \sum_{r \in \bZ + \nu } \frac{ \psi^{\mu}_{r} }{z^{r } }, ~~~~
\rho^A (z) = \sum_{r \in \bZ + \nu } \frac{ \rho^{A}_{r} }{z^{r }} .
\end{eqnarray}
The mode expansions for the DD(NN) directions and those for the ND(DN) directions differ in the power of expansions by one half.  
Also, the zero point energy is changed by $\frac{1}{8} ( \nu_B - \nu_F ) $ only in the NS sector.  
By use of these, we compute the amplitude $F$ between one ${\rm D}_{p|r}$-brane and one ${\rm D}_{q|\tilde{r}}$-brane.  
\begin{equation}
F = \int_0^{\infty} \frac{d \tau}{2 \tau} {\rm Tr} ~ e^{- 2 \pi \tau L_0}
  = \int_{0}^{\infty} \frac{d \tau}{2 \tau} ~ Z_{X \Theta} Z_{\psi \rho} \label{06}
\end{equation}
Here the Virasoro generator $L_0$ is 
\begin{eqnarray}
L_0 &=& {\alpha}' (k^{\alpha} k_{\alpha} - i w^a w_a ) + \frac{y^2}{4 \pi^2 {\alpha}'} 
+ \sum_{n=\bZ+ \hat{\nu}} ( \alpha^{\mu}_{-n} \alpha_{\mu n} - i \beta^A_{-n} \beta_{An}  ) \nonumber \\
& & + \sum_{r \in \bZ + \nu + \hat{\nu} } r ( \psi^{\mu}_{-r} \psi_{\mu r} - i \rho^A_{-r} \rho_{A r} ) 
+ L_0^{\rm ghost} .
\end{eqnarray}
The index $\hat{\nu}$ means that $\hat{\nu}=\frac{1}{2}$ for the ND(DN) directions and  $\hat{\nu}=0$ for the others.  The part of $X^{\mu}$ and $ \Theta^A$ is 
\begin{equation}
Z_{X \Theta} = i ( 8 \pi^2 {\alpha}' \tau )^{- \frac{NN_B}{2} + \frac{NN_F}{2}}  ~ e^{- \frac{y^2 \tau}{2 \pi {\alpha}'}}
       ~ f_1^{-(\mu_B - \mu_F)} ~ f_4^{- ( \nu_B-\nu_F)} .
\end{equation}
The part of $\psi^{\mu}$ and $\rho^A$ is 
\begin{equation}
Z_{\psi \rho} = \frac{1}{2} \left[ f_3^{\mu_B - \mu_F} f_2^{\nu_B - \nu_F} 
         - f_4^{\mu_B - \mu_F} \delta_{\nu_B - \nu_F,0} - f_2^{\mu_B - \mu_F} f_3^{\nu_B - \nu_F} 
  - f_4^{\nu_B - \nu_F} \delta_{\mu_B - \mu_F,0} \right] .
\end{equation}
Here we define $\mu_B=D_B - \nu_B - 2$ and $\mu_F = D_F - \nu_F$.  
The functions $f_i$ for $i =1,2,3,4$ are defined as
\begin{eqnarray}
f_1 = q^{\frac{1}{12}} \prod^{\infty}_{n=1} (1 - q^{2n}), ~~& &~~f_2 = \sqrt{2} ~ q^{\frac{1}{12}} \prod^{\infty}_{n=1} (1 + q^{2n}) , \\
f_3 = q^{-\frac{1}{24}} \prod^{\infty}_{n=1} (1 + q^{2n-1}), ~~& & ~~ f_4 = q^{- \frac{1}{24}} \prod^{\infty}_{n=1} (1-q^{2n-1}) .
\end{eqnarray}
After substituting these into (\ref{06}), the amplitude is 
\begin{eqnarray}
F &=& i ( 8 \pi^2 {\alpha}' )^{- \frac{NN_B}{2}+ \frac{NN_F}{2}}  \int_{0}^{\infty} 
\frac{d \tau}{\tau} ~ {\tau}^{- \frac{NN_B - NN_F}{2}} e^{- \frac{y^2 \tau}{2 \pi {\alpha}'}} \nonumber  \\
 & & \times \frac{1}{2} \left[ \left( \frac{f_3}{f_1} \right)^{\mu_B - \mu_F} \left( \frac{f_2}{f_4} \right)^{\nu_B - \nu_F} 
- \left( \frac{f_4}{f_1} \right)^{\mu_B - \mu_F}   \delta_{\nu_B - \nu_F,0} \right. \nonumber \\
  & & \left.  ~~~~~~     - \left(\frac{f_2}{f_1} \right)^{\mu_B - \mu_F} \left( \frac{f_3}{f_4} \right)^{\nu_B - \nu_F} 
- \delta_{\mu_B - \mu_F,0} \right] .  \label{open amp}
\end{eqnarray}
Here the last term exists only in the case that $\nu_B - \nu_F=8$, because there are no fermionic zero modes in the $R(-1)^F$ sector 
and this sector contributes \cite{200}.  
The well-known abstruse identity is 
\begin{equation}
f_4^8 - f_3^8 - f_2^8 =0 . \label{011}
\end{equation}
After using this abstruse identity, we can see that the amplitude $F$ vanishes if $\nu_B-\nu_F=0,4$ and $8$.  
Therefore we find the configurations of two D-branes which satisfy the BPS-like no-force conditions.

Next we consider the closed string channel.  In order to construct the boundary states, the following conditions are imposed.
\begin{eqnarray}
\left( \psi^{\mu} - i \eta {S^{\mu}}_{ \nu} {\tilde{\psi}}^{\nu} \right) |_{\sigma_2=0} | B ,\eta\rangle =0 ,\\
\left( \rho^{A} - i \eta {S^{AB}} {\tilde{\rho}}_{B} \right) |_{\sigma_2=0} | B ,\eta \rangle =0.
\end{eqnarray}
Here we introduce the parameter $\eta= \pm 1$ on the boundary, $\sigma_2=0$, and the parameter $\tilde{\eta}= \pm 1$ on the boundary, $\sigma_2=\pi$.
%
By use of the oscillators, these conditions are rewritten as  
\begin{eqnarray}
\left( \psi^{\mu}_r - i \eta {S^{\mu}}_{ \nu} {\tilde{\psi}}^{\nu}_{-r} \right) |_{\sigma_2=0} | B ,\eta \rangle =0  , \label{bbc1} \\
\left( \rho^{A}_r - i \eta {S^{AB}} {\tilde{\rho}}_{B -r} \right) |_{\sigma_2=0} | B , \eta \rangle =0 .
\end{eqnarray}
Also the conditions for the ghost parts are 
\begin{eqnarray}
\left( \gamma_r + i \eta {\tilde{\gamma}}_{-r} \right) |_{\sigma_2=0} | B , \eta \rangle = 0 ,\\
\left( \beta_r + i \eta {\tilde{\beta}}_{-r} \right) |_{\sigma_2=0} | B , \eta \rangle = 0. \label{bbc2}
\end{eqnarray}
The boundary states can be written as
\begin{equation}
| B  \rangle =  \frac{T_{p,r}}{2} |B_X \rangle |B_{\Theta} \rangle  ( |B \rangle_{{\rm NS }} + |B \rangle_{{\rm R }})  .
\end{equation}
In the NS-NS sector in the $(- 1, -1)$ picture, the boundary states are
\begin{eqnarray}
|B_{\psi} ,\eta \rangle &=&  - i \prod^{\infty}_{r=\frac{1}{2}} \left( e^{i \eta \psi^{\mu}_{-r} {S_{\mu}}^{\nu} {\tilde{\psi}}_{\nu -r}} \right) |0 \rangle ,\\
|B_{\rho} ,\eta  \rangle &=& \prod^{\infty}_{r=\frac{1}{2}} \left( e^{ \eta \rho^{A}_{-r} {S_{A}}^{B} {\tilde{\rho}}_{B -r}} \right) |0 \rangle ,\\
|B_{\rm ghost} ,\eta  \rangle  &=& \prod^{\infty}_{r=\frac{1}{2}} \left( e^{i \eta (\gamma_{-r} {\tilde{\beta}}_{-r} - \beta_{-r } {\tilde{\gamma}}_{-r})} \right) 
|P=-1 , \tilde{P} = -1 \rangle ,
\end{eqnarray}
where $|P , \tilde{P}  \rangle$ is defined as follows \cite{300}.
\begin{eqnarray}
& & \gamma_{m} |P , \tilde{P}  \rangle = 0 ~~~~ {\rm for } ~~ m \ge P + \frac{3}{2} ,\nonumber \\
& & \beta_{m} |P , \tilde{P}  \rangle = 0 ~~~~ {\rm for } ~~ m \ge -P - \frac{1}{2} .
\end{eqnarray}
Here $\gamma_{m}$ and $\beta_{m}$ are the oscillators for the superghosts $\gamma$ and $\beta$.  
Also, the conditions for $\tilde{\gamma}$ and $\tilde{\beta}$ are defined similarly.  
In the R-R sector in the $(- \frac{1}{2},- \frac{3}{2})$ picture, the boundary states are 
\begin{eqnarray}
|B_{\psi} ,\eta \rangle &=&  - \prod^{\infty}_{r=1} \left( e^{i \eta \psi^{\mu}_{-r} {S_{\mu}}^{\nu} {\tilde{\psi}}_{\nu -r}} \right) |B^{(0)}_{\psi} , \eta \rangle ,\\
|B_{\rho} ,\eta \rangle &=&  \prod^{\infty}_{r=1} \left( e^{ \eta \rho^{A}_{-r} {S_{A}}^{B} {\tilde{\rho}}_{B -r}} \right) |B^{(0)}_{\rho} ,\eta \rangle ,\\
|B_{\rm ghost}  ,\eta \rangle  &=& \prod^{\infty}_{r=1} \left( e^{i \eta (\gamma_{-r} {\tilde{\beta}}_{-r} - \beta_{-r } {\tilde{\gamma}}_{-r})} \right) 
|B^{(0)}_{\rm ghost} ,\eta \rangle .
\end{eqnarray}
The parts of the R-R zero modes are 
\begin{eqnarray}
|B^{(0)} ,\eta \rangle &=& |B^{(0)}_{\psi} ,\eta \rangle |B^{(0)}_{\rho} ,\eta \rangle |B^{(0)}_{\rm ghost},\eta \rangle ,\\
|B^{(0)}_{\psi} ,\eta \rangle &=& \left( C \Gamma^0 \cdots \Gamma^{p} \left( \frac{1+ i  \eta \Gamma_{D_B +1} }{1 + i \eta } \right) \right)_{A \tilde{B}} 
|A \rangle | \tilde{B} \rangle ,\\ 
|B^{(0)}_{\rho} ,\eta \rangle &=& i^{- \frac{r}{2}} \prod_{a = 1}^{\frac{r}{2}} e^{ \eta \rho^{2a-1}_0 {\tilde{\rho}}^{2a}_0}  
\prod^{\frac{D_F-r}{2}}_{s=\frac{r}{2}+1} e^{- \eta \rho^{2s-1}_0 {\tilde{\rho}}_0^{2s}} | 0 \rangle_{\rho}      ,\\
|B^{(0)}_{\rm ghost} ,\eta \rangle &=& e^{i  \eta \gamma_0 {\tilde{\beta}}_0 } | P=- \frac{1}{2} , \tilde{P}=- \frac{3}{2} \rangle .
\end{eqnarray}
Here the ground state $| 0 \rangle_{\rho}$ is defined by 
\begin{equation}
\rho_0^{2a-1} | 0 \rangle_{\rho}={\tilde{\rho}}_0^{2a} | 0 \rangle_{\rho}= 
\rho_0^{2s-1} | 0 \rangle_{\rho}={\tilde{\rho}}_0^{2s} | 0 \rangle_{\rho}= 0.  
\end{equation}
The $\gamma$-matrices of $SO(1,D_B-1)$ are represented as 
\begin{eqnarray}
\Gamma^{i} &=& \left( 
              \begin{array}{cc} 
              0 & \gamma^i \\
               \gamma^i & 0  \end{array}
                   \right)        ,\\
\Gamma^{D_B -1} &=& i^{- \frac{D_B -2}{2}} \left( 
              \begin{array}{cc} 
              0 & \gamma^1 \cdots \gamma^{D_B-2}  \\
               \gamma^1 \cdots \gamma^{D_B-2} & 0  \end{array}
                   \right)        ,\\
\Gamma^{0} &=& \left( 
              \begin{array}{cc} 
              0 & 1 \\
               -1 & 0  \end{array}
                   \right)        ,\\
\Gamma_{D_B+1} &=& i^{- \frac{D_B -2}{2}} \Gamma^0 \Gamma^1 \cdots \Gamma^{D_B-1} .
\end{eqnarray}
Here $\gamma^i$ is the $\gamma$-matrices of $SO(D_B-2)$, where $i=1,2,\cdots,D_B-2$ .  
The actions of the Ramond oscillators ${\psi}^{\mu}_{r}$ and ${\tilde{\psi}}^{\mu}_{r}$ on the state $|A \rangle | \tilde{B} \rangle$ are given by 
\begin{eqnarray}
& & \psi^{\mu}_0  |A \rangle | \tilde{B} \rangle = \frac{1}{\sqrt{2}} {(\Gamma^{\mu})^A}_C {( 1 )^B}_{D} |C \rangle | \tilde{D} \rangle \nonumber ,\\
& & {\tilde{\psi}}^{\mu}_0  |A \rangle | \tilde{B} \rangle = \frac{1}{\sqrt{2}} {(\Gamma_{D_B+1})^A}_C { (\Gamma^{\mu} )^B}_{D} |C \rangle | \tilde{D} \rangle \nonumber ,\\
& & \psi^{\mu}_r  |A \rangle | \tilde{B} \rangle = {\tilde{\psi}}^{\mu}_r  |A \rangle | \tilde{B} \rangle = 0 ~~~~ {\rm for} ~~ r \ge 1 .  
\end{eqnarray}
It is easy to check that these boundary states satisfy the boundary conditions from (\ref{bbc1}) to (\ref{bbc2}).  
The GSO projections for the NS-NS boundary states are given by
\begin{eqnarray}
|B \rangle_{NS}  &=& \frac{1 + (-1)^{F+ {F_{\rho} + G} }}{2} ~ \frac{1 + (-1)^{\tilde{F} + {\tilde{F}}_{\tilde{\rho}}+ \tilde{G}}}{2} 
~ |B_{\psi}, + \rangle |B_{\rho} , + \rangle |B_{\rm ghost} , + \rangle \nonumber \\
&=& \frac{1}{2} \Big[ |B_{\psi}, + \rangle |B_{\rho} , + \rangle |B_{\rm ghost} , + \rangle 
    - |B_{\psi}, - \rangle |B_{\rho} , - \rangle |B_{\rm ghost} , - \rangle  \Big] ,
\end{eqnarray}
where $F$, $F_{\rho}$ and $G$ are 
\begin{eqnarray}
F &=& \sum^{\infty}_{m=\frac{1}{2}} \psi^{\mu}_{-m} \psi_{\mu m} -1 ,\\
F_{\rho} &=&   - i \sum^{\infty}_{m=\frac{1}{2}} \rho^A_{-m} \rho_{A m} , \\
G &=& - \sum^{\infty}_{m=\frac{1}{2}} (\gamma_{-m} \beta_m + \beta_{-m} \gamma_{m}) ,
\end{eqnarray}
and $\tilde{F}$, ${\tilde{F}}_{\tilde{\rho}}$ and $\tilde{G}$ are defined similarly.  
The GSO projection for the R-R boundary states are given by
\begin{eqnarray}
|B \rangle_{R} &=& \frac{1 +  (-1)^{F+ F_{\rho} + G} }{2} ~ \frac{ 1- (-1)^{\frac{D_B-2}{2}} (-1)^p (-1)^{\tilde{F} + {\tilde{F}}_{\tilde{\rho}} + \tilde{G}}}{2} 
 ~   |B_{\psi}, + \rangle |B_{\rho} , + \rangle |B_{\rm ghost} , + \rangle                      \nonumber \\
&=& \frac{1}{2} \Big[ |B_{\psi}, + \rangle |B_{\rho} , + \rangle |B_{\rm ghost} , + \rangle 
    - (-1)^p |B_{\psi}, - \rangle |B_{\rho} , - \rangle |B_{\rm ghost} , - \rangle  \Big] ,
\end{eqnarray}
where $F$, $F_{\rho}$ and $G$ are defined by 
\begin{eqnarray}
(-1)^F &=& i^{- \frac{D_B-2}{2}} \psi^0_0 \psi^1_0 \cdots \psi^{D_B-1}_0 (-1)^{\sum_{m=1}^{\infty} \psi^{\mu}_{-m} \psi_{\mu m} -1} \\
F_{\rho} &=&   i \sum_{l=1}^{\frac{D_F}{2}} \rho_0^{2l} \rho_0^{2l-1} - i \sum^{\infty}_{m=1} \rho^A_{-m} \rho_{A m} \\
G &=& - \gamma_0 \beta_0 - \sum_{m=1}^{\infty} ( \gamma_{-m} \beta_m + \beta_{-m} \gamma_{m} ) 
\end{eqnarray}
and $\tilde{F}$, ${\tilde{F}}_{\tilde{\rho}}$ and $\tilde{G}$ are defined similarly.  
Here $p$ is an even number in type IIA and $p$ is an odd number in type IIB, when $\frac{D_B-2}{2}$ is an even number.  
Also, $p$ is an odd number in type IIA and $p$ is an even number in type IIB, when $\frac{D_B-2}{2}$ is an odd number.

We compute the static interaction between one $D_{p|r}$-brane localized at $x^i=0$ and one $D_{q|\tilde{r}}$ brane localized at $x^i=y^i$.  
%
\begin{equation}
F = \langle B_1 | D | B_2 \rangle  = \frac{T_{p,r} T_{q,\tilde{r}} }{4} \frac{{\alpha}'}{4 \pi} \int_{1 \ge |z|} \frac{d^2 z}{|z|^2} ~
Z_{X} Z_{\Theta} ( Z_{{\rm NS}} + Z_{{\rm R}} ) . \label{2000}
\end{equation}
The parts of $X^{\mu}$ and $\Theta^{A}$ are 
\begin{eqnarray}
Z_{X} &=& i ( 2 \pi^2 {\alpha}' t )^{- \frac{DD_B }{2}} e^{ - \frac{y^2}{2\pi {\alpha}' t}}  
\prod^{\infty}_{n=1} \left( \frac{1}{1 - |z|^{2n}} \right)^{D_B - \nu_B} \left( \frac{1}{1 + |z|^{2n}} \right)^{\nu_B}  ,\\
Z_{\Theta} &=& (2 \pi^2 {\alpha}' t )^{\frac{DD_F}{2}} \prod^{\infty}_{n=1} (1 - |z|^{2 n})^{D_F-\nu_F} (1 + |z|^{2 n})^{\nu_F} .
\end{eqnarray}
In the NS-NS sector, $Z_{{\rm NS}}$ is  
\begin{equation}
Z_{{\rm NS}} = \frac{1}{4} {}_{NS} \langle B | z^{N} {\bar{z}}^{\tilde{N}}  |B \rangle_{NS} = \sum_{\eta=\pm 1} \sum_{\tilde{\eta}=\pm 1}  (-1)^{ \frac{\eta \tilde{\eta}-1}{2}}      
~ Z^{NS}_{\psi} (\eta,\tilde{\eta}) ~ Z^{NS}_{\rho} (  \eta,  \tilde{\eta}) ~ Z^{NS}_{{\rm Ghost}} (\eta,\tilde{\eta}) .
\end{equation}
The part of $\psi$ is 
\begin{eqnarray}
Z^{NS}_{\psi} (\eta ,\tilde{\eta}) &=& \langle B_{\psi} , \tilde{\eta} | z^{N} {\bar{z}}^{\tilde{N}} |B_{\psi} , \eta \rangle \nonumber \\
&=& \langle 0 | \left( \prod^{\infty}_{s=\frac{1}{2}}  e^{i \tilde{\eta} \psi^{\mu}_{s} {S_{\mu}}^{\nu} {\tilde{\psi}}_{\nu s}} \right)
z^{N} {\bar{z}}^{\tilde{N}}  \left( \prod^{\infty}_{r=\frac{1}{2}}  e^{i \eta \psi^{\mu}_{-r} {S_{\mu}}^{\nu} {\tilde{\psi}}_{\nu -r}} \right) |0 \rangle \nonumber \\
&=& \prod^{\infty}_{n=1} \left( 1 + \eta \tilde{\eta} ~  |z|^{2n-1} \right)^{D_B-\nu_B} \left( 1 - \eta \tilde{\eta} ~ |z|^{2n-1} \right)^{\nu_B} .
\end{eqnarray}
The part of $\rho$ is 
\begin{eqnarray}
 Z^{NS}_{\rho} (\eta , \tilde{\eta}) &=& \langle B_{\rho} , \tilde{\eta} | z^{N} {\bar{z}}^{\tilde{N}} |B_{\rho} , \eta \rangle \nonumber \\ 
&=& \langle 0 | \left( \prod^{\infty}_{s=\frac{1}{2}}  e^{ \tilde{\eta} \rho^{A}_{s} {S_{A}}^{B} {\tilde{\rho}}_{B s}} \right) 
z^{N} {\bar{z}}^{\tilde{N}} 
\left( \prod^{\infty}_{r=\frac{1}{2}}  e^{ \eta \rho^{A}_{-r} {S_{A}}^{B} {\tilde{\rho}}_{B -r}} \right) |0 \rangle \nonumber \\
&=& \prod^{\infty}_{n=1} \left( \frac{1}{1 +  \eta \tilde{\eta} ~ |z|^{2n-1}} \right)^{D_F-\nu_F} \left( \frac{1}{1 - \eta \tilde{\eta} ~ |z|^{2n-1} } \right)^{\nu_F} .
\end{eqnarray}
The ghost part is 
\begin{equation}
Z^{NS}_{{\rm Ghost} }(\eta , \tilde{\eta})  = |z|^{-1} \prod^{\infty}_{n=1} \left(  1- |z|^{2n} \right)^{2}  \left( \frac{1}{1 + \eta \tilde{\eta} ~ |z|^{2n-1}} \right)^2  .
\end{equation}
After substituting these parts into (\ref{2000}), the NS-NS amplitude is  
\begin{eqnarray}
A_{{\rm NS}-{\rm NS}} &=& \frac{T_{p,r} T_{q,\tilde{r}} }{4} \frac{{\alpha}'}{4 \pi} \int_{1 \ge |z|} \frac{d^2 z}{|z|^2} ~ 
Z_{X} Z_{\Theta}  Z_{{\rm NS}}
\nonumber \\
  &=& \frac{T_{p,r} T_{q,\tilde{r}}}{8} (i {\alpha}' \pi ) \int_{0}^{\infty} d t ~ e^{\pi t} ~ e^{- \frac{y^2}{2 \pi {\alpha}' t} } 
~ (2 \pi^2 {\alpha}' t )^{-\frac{DD_B}{2} + \frac{DD_F}{2}}  \nonumber \\
& & \times \frac{1}{2} \prod^{\infty}_{n=1} \left[ \left( \frac{1 +e^{- (2n-1) \pi t}}{1 - e^{- 2n \pi t}} \right)^{D_B-D_F-2 -\nu_B+\nu_F} 
\left( \frac{1  - e^{- (2n-1) \pi t}}{1 + e^{- 2n \pi t}} \right)^{\nu_B - \nu_F} \right. \nonumber \\
& & \left. - \left( \frac{1  - e^{- (2n-1) \pi t}}{1 - e^{- 2n \pi t}} \right)^{D_B-D_F-2 -\nu_B+\nu_F} 
\left( \frac{1 + e^{- (2n-1) \pi t}}{1 + e^{- 2n \pi t}} \right)^{\nu_B - \nu_F} \right] \nonumber  \\
&=& K \int^{\infty}_{0} dt ~ t^{- \frac{DD_B-DD_F}{2}} ~ e^{- \frac{y^2}{2 {\alpha}' \pi t }}  \nonumber \\
& & \times \frac{1}{2} \left[ \left( \frac{f_3}{f_1} \right)^{\mu_B - \mu_F} \left( \frac{f_4}{f_2} \right)^{ \nu_B - \nu_F}  
- \left( \frac{f_4}{f_1} \right)^{\mu_B - \mu_F} \left( \frac{f_3}{f_2} \right)^{ \nu_B - \nu_F} \right] \label{98} ,
\end{eqnarray}
where the coefficient $K$ is
\begin{eqnarray}
K &=& \frac{T_{p,r} T_{q,\tilde{r}}}{8} (i {\alpha}' \pi ) (2 \pi^2 {\alpha}'  )^{-\frac{DD_B}{2} + \frac{DD_F}{2}} (\sqrt{2})^{\nu_B - \nu_F} \nonumber \\
 &=& i (8 \pi^2 {\alpha}')^{- \frac{NN_B}{2} + \frac{NN_F}{2}} .
\end{eqnarray}

Next, in the R-R sector, $Z_{{\rm R}}$ is  
\begin{equation}
Z_{{\rm R}} = \frac{1}{4} {}_{R} \langle B | z^{N} {\bar{z}}^{\tilde{N}}  |B \rangle_{R} = \sum_{\eta=\pm 1} \sum_{\tilde{\eta}=\pm 1} \left[ - (-1)^p \right]^{\frac{1 - \eta \tilde{\eta}}{2}}      
Z^R_{\psi} (\eta,\tilde{\eta}) ~ Z^R_{\rho} (  \eta,  \tilde{\eta}) ~ Z^R_{{\rm Ghost}} (\eta,\tilde{\eta}) . \label{2001}
\end{equation}
The part of $\psi$ is
\begin{eqnarray}
Z^R_{\psi} (\eta, \tilde{\eta}) &=& |z|^{\frac{D_B}{8}} \langle B_{\psi}^{(0)} ,\tilde{\eta} | \left( \prod^{\infty}_{s=1}  
e^{i \tilde{\eta} \psi^{\mu}_{s} {S_{\mu}}^{\nu} {\tilde{\psi}}_{\nu s}} \right)
z^{N} {\bar{z}}^{\tilde{N}} \left( \prod^{\infty}_{r=1}  
e^{i \eta \psi^{\mu}_{-r} {S_{\mu}}^{\nu} {\tilde{\psi}}_{\nu -r}} \right) |B_{\psi}^{(0)} , \eta \rangle  \nonumber \\
&=& |z|^{\frac{D_B}{8}} \prod^{\infty}_{n=1}  \left( 1 + \eta \tilde{\eta} |z|^{2n} \right)^{D_B-\nu_B} 
\left( 1 - \eta \tilde{\eta} |z|^{2n} \right)^{\nu_B} \langle B_{\psi}^{(0)}, \tilde{\eta} |B_{\psi}^{(0)} ,\eta \rangle  .
\end{eqnarray}
The part of $\rho$ is 
\begin{eqnarray}
 Z^R_{\rho} (\eta , \tilde{\eta}) &=&  |z|^{- \frac{D_F}{8}} \langle B_{\rho}^{(0)} ,\tilde{\eta} | \left( \prod^{\infty}_{r=1}  e^{ \tilde{\eta} \rho^{A}_{r} {S_{A}}^{B} {\tilde{\rho}}_{B r}} \right) 
z^{N} {\bar{z}}^{\tilde{N}} 
\left( \prod^{\infty}_{r=1}  e^{ \eta \rho^{A}_{-r} {S_{A}}^{B} {\tilde{\rho}}_{B -r}} \right) |B_{\rho}^{(0)} , \eta \rangle \nonumber \\
&=& |z|^{- \frac{D_F}{8}} \left( \frac{1}{1 + \eta \tilde{\eta} |z|^{2n}} \right)^{D_F-\nu_F} \left( \frac{1}{1 - \eta \tilde{\eta } |z|^{2n} } \right)^{\nu_F} 
\langle B_{\rho}^{(0)} ,\tilde{\eta} | B_{\rho}^{(0)} , \eta \rangle  .
\end{eqnarray}
The part of the ghost is 
\begin{equation}
Z^R_{{\rm Ghost} } = |z|^{- \frac{5}{4}}  \prod^{\infty}_{n=1} \left(  1- |z|^{2n}  \right)^{2}
\left( \frac{1}{1 + \eta \tilde{\eta} |z|^{2n}} \right)^2 \langle B_{{\rm ghost}}^{(0)} , \tilde{\eta} |B_{{\rm ghost}}^{(0)} ,\eta  \rangle .
\end{equation}
After substituting these parts into (\ref{2001}), the R-R amplitude is  
\begin{eqnarray}
A_{{\rm R}-{\rm R}} &=&  \frac{T_{p,r} T_{q,\tilde{r}} }{4} \frac{{\alpha}'}{4 \pi} \int_{1 \ge |z|} \frac{d^2 z}{|z|^2} ~ 
~ Z_{X} Z_{\Theta}  Z_{{\rm R}} \nonumber \\
  &=& \frac{T_{p,r} T_{q,\tilde{r}}}{8} (i {\alpha}' \pi ) \int_{0}^{\infty} d t ~   e^{- \frac{y^2}{2 \pi {\alpha}' t} } 
~ (2 \pi^2 {\alpha}' t )^{-\frac{DD_B}{2} + \frac{DD_F}{2}} \nonumber \\
& & \times \frac{1}{2} \prod^{\infty}_{n=1} \left[ 
\left( \frac{1  + e^{- 2n \pi t}}{1 - e^{- 2n \pi t}} \right)^{D_B-D_F-2 -\nu_B+\nu_F} 
\left( \frac{1 -e^{-2n \pi t}}{1 + e^{-2n \pi t}} \right)^{\nu_B - \nu_F}  \delta_{\eta \tilde{\eta} , 1} 
\right. \nonumber \\
& & \left. - (-1)^p 
\left( \frac{1 - e^{-2n \pi t}}{1 - e^{- 2n \pi t}} \right)^{D_B-D_F-2 -\nu_B+\nu_F} 
\left( \frac{1  + e^{-2n \pi t}}{1 + e^{-2n \pi t}} \right)^{\nu_B - \nu_F} \delta_{\eta \tilde{\eta} , -1} 
\right] \langle B^{(0)} ,\tilde{\eta} | B^{(0)} , \eta \rangle \nonumber  \\
&=& K ~ 2^{- \frac{\nu_B-\nu_F}{2}} \int^{\infty}_{0} dt ~ t^{- \frac{DD_B-DD_F}{2}} ~ e^{- \frac{y^2}{2 {\alpha}' \pi t }}  \nonumber \\
& & \times \frac{1}{2} \left[  \delta_{\eta \tilde{\eta} ,1} ~ 2^{- \frac{ \mu_B - \mu_F -\nu_B+\nu_F }{2}}
 \left( \frac{f_2}{f_1} \right)^{\mu_B - \mu_F }  - (-1)^p \delta_{\eta \tilde{\eta} ,-1}    \right] \langle B^{(0)}, \tilde{\eta} | B^{(0)} ,\eta \rangle  .
\end{eqnarray}
Here the part of zero mode is computed by use of the regularization \cite{9};
\begin{eqnarray}
\langle B^{(0)} ,\tilde{\eta}  | B^{(0)} , {\eta} \rangle &=&  \lim_{x \rightarrow 1 } ~ 
\langle B_{\psi}^{(0)}, \tilde{\eta}| x^{2 F} |B_{\psi}^{(0)} ,\eta \rangle ~ 
\langle B_{\rho}^{(0)} , \tilde{\eta} | x^{2 F_{\rho}}  | B_{\rho}^{(0)} ,  \eta \rangle ~ 
\langle B_{{\rm ghost}}^{(0)} , \tilde{\eta} | x^{2 G} | B_{{\rm ghost}}^{(0)} ,\eta  \rangle \nonumber \\
&=& \lim_{x \rightarrow 1 } x^{- \frac{D_B}{2}} i^{\frac{\nu_B -\nu_F}{2}} \left[ - \delta_{\eta \tilde{\eta}, 1} ~ 
\left( x + \frac{1}{x} \right)^{\frac{D_B - \nu_B}{2}} \left( x - \frac{1}{x} \right)^{\frac{\nu_B}{2}} \right. \nonumber \\
& & ~~~~~~~~~ ~~~ ~~~~~~ \left. 
+ (-1)^p \delta_{\eta \tilde{\eta},-1} \left( x - \frac{1}{x} \right)^{\frac{D_B - \nu_B}{2}} \left( x + \frac{1}{x} \right)^{\frac{\nu_B}{2}} \right] \nonumber \\
& & ~~~~~~~~~~~~~~ \times \left( \frac{x^2}{x^2 + \eta \tilde{\eta} } \right)^{\frac{D_F - \nu_F} {2} } 
\left( \frac{x^2}{x^2 - \eta \tilde{\eta} } \right)^{\frac{\nu_F} {2} } ~ 
\frac{1}{1 + \eta \tilde{\eta} ~ x^2} \nonumber \\
&=&  - 16 ~ \delta_{\nu_B - \nu_F,0} ~ \delta_{\eta \tilde{\eta} ,1} 
+ 16 (-1)^{p} ~ \delta_{\nu_B - \nu_F , 8} ~ \delta_{\eta \tilde{\eta} , -1}  ,
\end{eqnarray}
where $0 \le \nu_B - \nu_F \le 8$.
The result is 
\begin{eqnarray}
A_{R-R} &=& K \int^{\infty}_{0} dt ~ t^{- \frac{DD_B-DD_F}{2}} ~ e^{- \frac{y^2}{2 {\alpha}' \pi t }} \nonumber \\
& & \times \frac{1}{2} \left[
- \left( \frac{f_2}{f_1} \right)^{\mu_B- \mu_F} \delta_{\nu_B - \nu_F,0} - \delta_{\nu_B - \nu_F,8}   \right] \label{99} .
\end{eqnarray}
After summing up the NS-NS part (\ref{98}) and the R-R part (\ref{99}), we can see that the total amplitude is
\begin{eqnarray}
A &=& A_{NS-NS} + A_{R-R} \nonumber \\
&=& K \int^{\infty}_{0} dt ~ t^{- \frac{DD_B-DD_F}{2}} ~ e^{- \frac{y^2}{2 {\alpha}' \pi t }}  \nonumber \\
& & \times \frac{1}{2} \left[ \left( \frac{f_3}{f_1} \right)^{\mu_B - \mu_F} \left( \frac{f_4}{f_2} \right)^{ \nu_B - \nu_F}  
- \left( \frac{f_4}{f_1} \right)^{\mu_B - \mu_F} \left( \frac{f_3}{f_2} \right)^{ \nu_B - \nu_F} \right. \nonumber \\
& & ~~~~~~  \left. - \left( \frac{f_2}{f_1} \right)^{\mu_B- \mu_F} \delta_{\nu_B - \nu_F,0} -  \delta_{\nu_B - \nu_F,8}   \right] .  \label{closed amp}
\end{eqnarray}
Also, these functions $f_i$ have the well-known properties
\begin{eqnarray} 
f_1 (e^{- \frac{\pi}{t}}) &=& \sqrt{t} f_1 ( e^{- \pi t}) ,\\
f_2 (e^{- \frac{\pi}{t}}) &=&  f_4 ( e^{- \pi t}) ,\\
f_3 (e^{- \frac{\pi}{t}}) &=&  f_3 ( e^{- \pi t}) .
\end{eqnarray}
It is easy to see that each term of (\ref{open amp}) is related to each term of (\ref{closed amp}) from these properties of $f_i$.  
In conclusion, the amplitude in the closed string channel is completely the same as that in the open string channel.  
Thus, we show that there is no net force between two D-branes if $\nu_B - \nu_F=0,4$ and $8$.  

%
~ \\
~ \\

\section{\Large Conclusion and Discussion}


In this paper, we have discussed bosonic string theories, RNS string theories and heterotic string theories on flat supermanifolds.  
We have shown that the central charges vanish for bosonic string theories on supermanifolds and they are modular invariant.  
Furthermore, we have constructed type II, type 0, type I string theories and heterotic string theories on supermanifolds.  
In the RNS string theories on supermanifolds, the fields $\rho^A$ have been treated as the $\beta \gamma$ superghosts with the weights $(\frac{1}{2},\frac{1}{2})$.  
We have shown that the one-loop vacuum amplitudes of type II, type I string theories and heterotic string theories are exactly zero .  
This may imply the existence of supersymmetry on supermanifolds.  
A surprising point is that the spinor representation for the supergroup is an infinite dimensional representation.  
It would be interesting to study supersymmetry on supermanifolds more.  
It would also be important to study the field theory on supermanifolds to check anomaly cancellations.  
It would also be interesting to find the relation between supersymmetry on supermanifolds and other physical theories such as the noncommutative geometry, 
which is related to the Heisenberg algebra as the commutators of the coordinates.

Although twistor string theory is non-unitary \cite{5}, it has given a new important viewpoint about the amplitudes of Yang-Mills theory in 4 dimensional space.  
Similarly in the case of the string theories on supermanifolds discussed in this paper, it would also be interesting to study how to restore the unitarity by using the property of string theory such as tachyons and orbifolding.  
Here we naively discuss one possibility to obtain the relation between the bosonic string theories on supermanifolds and well-known unitary theories.  
In the case of bosonic string theories on supermanifolds, the effective field theories on target spaces are the Yang-Mills theory and the Einstein gravity theory on supermanifolds.  
Long time ago, the Einstein gravity theory on supermanifolds was studied as a candidate for the supergravity theory by Arnowitt and Nath \cite{41}.  
Also, they tried to find the relation between the Einstein gravity theory on supermanifolds and the well-known supergravity theory.  
The well-known supergravity theory is the gravity theory on superspace.  
But it is still unclear to understand the relation between the gravity theory on a supermanifold and that on a superspace.  
Here we discuss the difference between supermanifolds and superspaces.  
The group acting on a flat supermanifold is the Poincare group whose labels run over both bosonic and fermionic 
coordinates.  Also, the group acting on a superspace is the supergroup whose algebra includes supersymmetric charges.  
These groups are actually different from each other, but we can see that the Poincare algebra ``includes'' the supersymmetric algebra after using the twisting and the In\"{o}nu-Wigner contraction \cite{42}.  
The twisting can be used to convert the labels of the fermionic coordinates to the spinor indices in a Minkowski space.  
If we understand the relation between supermanifolds and superspaces more clearly, we expect to be able to know the relation between string theories on supermanifolds and string theories on superspaces.  
The string theory on superspace is the well-known Green-Schwarz superstring theory, which has the $\kappa$ symmetry \cite{43}.  
In the case of the Green-Schwarz superstring theory, the $\kappa$ symmetry is used to truncate negative norm states.  
Thus, it would be interesting to find the relation between the bosonic string theories on supermanifolds and the Green-Schwarz superstring theories clearly in order to restore the unitarity.  
Also, the development of the hybrid and pure spinor formalisms by Berkovits would be important to know how to extract the physical unitary theories from these string theories on supermanifolds \cite{a3}.  

As mentioned in introduction, it is well-known that the partition functions of the topological string theories on bosonic manifolds are equivalent 
to the topological amplitudes of the type II superstring theory.  
Also, topological string theories on supermanifolds have been studied.  
Therefore, it would be very interesting to check whether the type II string theories on supermanifolds are related to topological string theories on supermanifolds.  

\section*{\Large Acknowledgments}
The author would like to thank our colleagues at the Yukawa Institute for Theoretical Physics and the members of the theoretical particle physics group at Kyoto university.  
The author is especially thankful to H. Kajiura, H. Kunitomo, H. Shimada and S. Sugimoto for useful discussions and encouragement.  
Also, the author is grateful to Y. Aisaka, K. Hasebe, S. Kawai, T. Kugo, T. Morita, T. Onogi, N. Sasakura, S. Seki and T. Takimi for valuable discussions and encouragement.

\end{document}